\documentclass[twocolumn,trackchanges,twocolappendix]{aastex631} 
\shorttitle{Modelling Streams and Tails}
\shortauthors{Pippert et al.}
\graphicspath{{./}{figures/}}
\usepackage{amsmath}
\usepackage{todonotes}
\usepackage{textcomp}
\usepackage{gensymb}
\usepackage{tikz}
\usepackage{rotating}
\usepackage{placeins}
\usepackage{subfigure}
\usepackage{orcidlink}

\begin{document}

\title{Modeling Tidal Streams and Tidal Tails Around Galaxies \\ Using Deep Wendelstein Imaging Data}

\author[0009-0006-9461-002X]{Jan-Niklas Pippert}
\thanks{jnpippert@mpe.mpg.de}
\affiliation{Max Planck Institute for Extraterrestrial Physics, Giessenbachstrasse, D-85748 Garching, Germany}
\affiliation{University Observatory, Faculty of Physics, Ludwig-Maximilians-Universität München, Scheinerstr. 1, 81679 Munich, Germany}

\author[0000-0002-9618-2552]{Matthias Kluge}
\affiliation{Max Planck Institute for Extraterrestrial Physics, Giessenbachstrasse, D-85748 Garching, Germany}
\affiliation{University Observatory, Faculty of Physics, Ludwig-Maximilians-Universität München, Scheinerstr. 1, 81679 Munich, Germany}

\author[0000-0001-7179-0626]{Ralf Bender}
\affiliation{Max Planck Institute for Extraterrestrial Physics, Giessenbachstrasse, D-85748 Garching, Germany}
\affiliation{University Observatory, Faculty of Physics, Ludwig-Maximilians-Universität München, Scheinerstr. 1, 81679 Munich, Germany}



\begin{abstract}
From the $\Lambda$ cold dark matter paradigm it is expected that galaxies merge and grow in their environments. These processes form various tidal features depending on the merger mass ratio, orbital parameters, and gas richness. We inspected 170 $g'$-band Abell cluster observations from the 2.1\,m Fraunhofer-Teleskop Wendelstein and identify 111 of such features from which we select nine streams and five tails. A fast and innovative technique was developed for determining their photometric properties. The model is a Gaussian, including higher-order moments to describe the light profile in slices perpendicular to the elongation direction. From these models, FWHM apertures are generated. The method was developed, tested, and applied on the selected features and corresponding $g$- and $r$-band data from the Legacy Survey DR10. Regarding the novel modeling approach, we can measure surface and total brightnesses with precisions of 4\% and 7\%, respectively. Mean stream width precision, which also translates to the mean $R_\mathrm{e}$ along the feature is on average within 3\% uncertainty. The measured streams have on average a surface brightness of $\sim26.25$ $g'$ mag arcsec$^{-2}$ and are dimmer than the tails in our sample ($\sim25.14$ $g'$ mag arcsec$^{-2}$). We infer that the progenitors of our streams can come from dwarfs, early-type galaxies or disks, based on the streams structural parameters. Furthermore, brightnesses and colors of the streams and tails are consistent with those of galaxies that populate the red sequence in the Coma cluster within $2\sigma$.
\end{abstract}



\section{Introduction} \label{sec:intro}
Structures in a cold dark matter Universe with a cosmological constant $\Lambda$ assemble hierarchically through the merging of smaller dark matter halos. Those potential wells bind baryonic matter, initiating star formation and creating galaxies, which contain dark matter, gas, and stars. From there, galaxies grow through the process of galaxy mergers, which dictates their morphological appearance and kinematic evolution.

Mergers are commonly classified into three categories by the mass ratios $\mu$ of the merging galaxies \citep[e.g.,][]{schulze+2020}, namely major mergers ($\mu\geq1/4$), minor mergers ($1/10<\mu<1/4$), and mini mergers ($\mu\leq1/10$). While major mergers are rarer, about three times at $z\sim 0.7$ \citep{lotz+2011} and about two times for $0<z<3$ \citep{conselice+2022}, accretion of small satellites is expected to be more common. This circumstance is also supported by the luminosity function \citep{schechter1976} of galaxies, where a significantly larger number of smaller galaxies is present. During the merger process, stellar material is ejected from the system, or stripped from an infalling satellite galaxy creating faint substructures around their host galaxies, in the form of streams, shells, tails, and continuous subforms. They store information of the past evolution of the system and give valuable insights into merger histories. The classification of these features in this work is based on the definitions used among the literature \citep[e.g.,][]{duc+2014,bilek+2020,sola+2022,urbano+2024}.

Tails originate in major galaxy merger events where the contents of both partners are ejected outward. These trails can extend up to 100 kpc \citep{toomre+1972}. As they come from their host, they show the same properties, e.g., age, metallicity, and color if no star formation was triggered by the merger. However, depending on the type of the merging galaxies, tails can be either red or show clumps of star formation and bluer colors \citep{elmegreen+2007} or a superposition of both \citep{mulia+2015}. Their usual appearance consists of elongated structures on each side of the galaxy, where their most prominent example is the Antennae galaxies \citep[e.g.,][]{lahen+2018}.

Streams, on the other hand, have a different history. They are the result of a low-mass accretion \citep{bullock+2005}, i.e., minor merger, where a progenitor gets accreted through dynamical friction and tidal forces and consequently stripped when entering the Roche limits, where the effect is strongest at the pericentric passage. The satellite infall for this definition of streams is biased toward circular infall \citep{karademir+2019}. While the stars disperse, they trace the orbit of the progenitor \citep[e.g.,][]{sanders+2013}, which can result in multiple loops around the host galaxy \citep[e.g.,][]{delgado+2008}. Streams can also occur in the process of forming shells, which is favored by radial infalls \citep{karademir+2019}.

Tidal streams in and around our Galaxy are detected either via star counts and groups of stars with coherent velocities \citep{malhan+2019,koposov+2023}, or groups of stars with the same chemical abundances \citep{martin+2022}. The stellar streams roughly follow the orbit of their destroyed satellite (galaxy or globular cluster), i.e., tracing the host potential, and are, hence, sensitive to the total matter within the orbit, including dark matter. Techniques such as orbital integration methods \citep[e.g.,][]{fardal+2006,koposov+2010,whelan+2014} and generative models for stream formation \citep[e.g.,][]{bonaca+2014,gibbons+2014} are developed to constrain halo potentials. A recent study by \citet{nibauer+2023} used another approach utilizing stream tracks and trial potentials to infer the flattening of the gravitational potential. An additional example is a study by \cite{ibata+2024}, which used $87$ streams to refine the mass model of the Milky Way. A unified collection of galactic streams and their footprints can be found in \cite{mateu+2023}.

Further examples of streams in the Local Group \citep{connachie+2018}, the Milky Way \citep{belokurov+2006,ibata+2021}, as well as in nearby galaxies, e.g., NGC 891 \citep{mouhcine+2010} and NGC 5128 \citep{crnojevic+2016}, are found by counting stars. Resolving stellar halos and their embedded substructures, e.g., streams, allows to reach deeper surface brightnesses than we can hope to reach with integrated photometry (e.g., due to galactic cirrus).

Due to their low surface brightness (LSB), imaging extragalactic tidal features is challenging but possible, as shown by surveys like MATLAS \citep{duc+2014,bilek+2020,sola+2022}, the Stellar Stream Legacy Survey \citep[SSLS;][]{delgado+2023b} and deep galaxy cluster observations \citep{kluge+2020}.

Some photometric studies of tidal streams by \cite{delgado+2008}, \cite{chonis+2011}, and \cite{roman+2023} exist, where surface brightnesses of individual examples ranging from $\sim26-28\,\,R$ mag arcsec$^{-2}$ and up to $\sim29.5\,\,g$ mag arcsec$^{-2}$ are measured. They obtained the surface brightness from the flux averaged inside one FWHM of a Gaussian distribution at either a collapsed representation of the stream or of a segment, which got extrapolated to the full extent of the feature. In addition, \cite{delgado+2021} analyzed the giant stream around M104 using the maximum surface brightness instead of the FWHM method. A few stream surveys already exist. For example, \cite{delgado+2023b} measured the surface brightnesses in the $g$, $r$, and $z$ bands for galaxies at redshift $z\lesssim0.02$, in many small circular apertures arranged along the feature. They reach down to $28.4\,\,g$ mag arcsec$^{-2}$, $26.9\,\,r$ mag arcsec$^{-2}$, and $26.5\,\,z$ mag arcsec$^{-2}$. Another example using the same method is the one by \citep{carretero+2023}, who provides $(g-r)$ colors and $g$- and $r$-band surface brightnesses down to $28.7$ and $27.98$ mag arcsec$^{-2}$ respectively.

Such studies can further be used to determine the stellar populations of the streams. Again, with streams in and around the Milky Way in which single stars are resolved, spectral classes, ages, and metallicities for each star can be obtained from the data directly. For those with an extragalactic origin, sufficiently deep spectra require long integration times and can be limited by systematics such as sky removal or continuum calibration. Therefore, one has to rely on flexible stellar population synthesis models \citep{laine+2016}, colors \citep{foster+2014}, integrated surface photometry \citep{delgado+2023a}, or SED fitting \citep{laine+2024}.

With the launch of the Euclid space telescope, on 2023 July 1 \citep[\textit{EUCLID};][]{euclid}, and upcoming deep optical surveys such as LSST with the Rubin Observatory \citep{ivezic+2019} or the Nancy Grace Roman Space Telescope \citep{akeson+2019}, a highly increased number of detected LSB structures is expected. Methods relying on manual inspection will not be feasible to statistically study the properties of tidal features in galaxy groups, galaxy clusters, or around field galaxies. In this work, we focus on streams and tails and present the first automated method to create robust and precise models from deep images. The models, based on Gaussian distributions, including higher-order Hermite basis functions, generate photometric apertures defined by an FWHM criterion. In such a manner, the brightness, colors, and shape parameters of streams and tails are obtained automatically.

Our novel modeling algorithm is a crucial ingredient for fully automated pipelines, which is useful for many applications as shown above. However, identifying these structures remains a manual task (in this work), but preparations \citep{sola+2022} and efforts to develop automated detection methods using neural networks \citep[e.g.,][]{richards2022,sanchez+2023,desmons+2024} are an active field of research.

In Section \ref{sec:data} we present the data used in this work. All steps and the methods involved in producing the stream and tail models and measuring brightnesses are described in Section \ref{sec:methods}. Stellar streams originate from disrupted or disturbed galaxies. In the case of total disruption and absence of star formation, their total brightnesses conserved. Their color of the stellar population moves slightly to redder colors. From this consideration, we expect that streams that originate from dust-free passive galaxies remain on the red sequence. The validity of this statement and whether streams originate from passive galaxies or form a different class is addressed and discussed in Section \ref{sec:results}. Tidal tails are only used to compare their colors to the streams and to test the modeling algorithm.

Throughout this work, we assume a flat cosmology with $H_0 = 69.6$ km s$^{-1}$ Mpc$^{-1}$ and $\Omega_m=0.286$ \citep{bennett+2014}. Luminosity distances and angular scales were calculated with the cosmological calculator by \cite{wright2006}. All magnitudes are given in the AB system. When referring to \textit{a feature}, the statement always includes both streams and tails. Otherwise, they are mentioned separately.

\section{Data}\label{sec:data}
\subsection{Telescopes and Surveys}
\subsubsection{2.1\,m Wendelstein Telescope}
Primarily, the 170 $g'$ band Abell galaxy cluster observations from \cite{kluge2020} are used in this work. All clusters fall into a redshift range of $z\lesssim0.08$. Additionally, for the A1656 (Coma Cluster), $r'$-band data was available. The data was captured with the 2.1\,m Fraunhofer Wendelstein telescope and its Wide Field Imager \citep[WWFI;][]{kosyra+2014}, which encompasses four ev2 CCDs, each with $4096\times4096$ 15\,$\mu$m pixels. Each detector has a field of view (FOV) of $13\farcm7\times13\farcm7$ and a pixel scale of $0\farcs2$ pixel$^{-1}$. A total FOV of $27\farcm6\times29\farcm0$ for one exposure is achieved which includes the $98\arcsec$ north-south and the $22\arcsec$ east-west gap. By applying a dither pattern of 52 steps, the final observed field is increased up to $\sim49\farcm3\times54\farcm0$.

\subsubsection{Legacy Survey - DECals and BASS}
To measure stream colors, we use existing WWFI $r'$ band information for features in the Coma cluster. For all other features, we utilize Legacy Survey DR10 data in the $g$ and $r$ bands. The survey is split into several subsurveys from which our WWFI fields were covered by the Beijing-Arizona Sky Survey (BASS) and the Dark Energy Camera Legacy Survey (DECaLS; \cite{dey+2019}). BASS uses the 2.3\,m Bok telescope equipped with the 90Prime wide field imager. It consists of four thinned Lockheed $4096\times4096$ pixel CCDs and has an FOV of $1.16\degree\times1.16\degree$ and a pixel scale of $0\farcs45$ pixel$^{-1}$. DECaLS uses the 4\,m Victor M. Blanco telescope. With 62 LBNL red-sensitive $2048\times4096$ pixel CCDs at the prime focus and a pixel scale of $0\farcs263$/pixel, it covers a field of 3 square degrees.

\subsection{Data Processing}
All data used was already reduced and coadded to final stacks. WWFI stacks were reduced with the WWFI data reduction pipeline \citep{kluge2020,kluge+2020,zoeller+2024}. It was developed for brightest cluster galaxy (BCG) and intracluster light studies, which hence includes precise background subtraction with night-sky flats. Zero-points are calibrated to the Pan-STARRS DVO PV3 catalog \citep{flewelling+2020} using apertures of $10''$ (50 pixels) diameter. Other effects such as additional offsets from the Pan-STARRS catalog, ghosts, and the missing light when increasing from $10''$ to $908''$ (infinity) apertures result in a zero-point correction of $0.1155$ magnitudes for the $g'$-band, hence ZP$_\text{inf} = $ZP$_{50} + 0.1155$ mag. When calculating $g'$ band brightnesses of the tidal features, ZP$_\text{inf}$ is used. However, for color analysis, ZP$_{50}$ is needed because $ZP_\text{inf}$ for the $r'$ is not determined yet.

We use the final reduced data from the LS DR10, which are processed by the NOIRLab Community and Legacy Pipeline. Cutouts of the features were automatically obtained using the LS Sky Viewer. For all cutouts, we used a pixel scale of $0\farcs263$ pixel$^{-1}$.

\subsection{Data Inspection}
The cluster stacks are manually inspected for signs of tidal features. We picked 15 features (nine streams and six tails), which have a high signal-to-noise ratio (S/N), show no sharp turns, i.e., a strong curvature, and are distinguishable from the merger/host. Further, these agree with the limitations of the modeling algorithm so we are sure that the modeling will succeed. We list those we found alongside their equatorial coordinates, that is the center of the feature, their redshift estimator, i.e., their host galaxy or cluster redshift, their angular scale, and their feature type in Table \ref{tab:feature-list}. All remaining tidal features, of any form, that were classified but not analyzed in this work are listed in Table \ref{tab:allfeatures} and Table \ref{tab:allfeaturescont}. We note that for some features the classification is not straightforward and that it might not even be a tidal feature at all.

\begin{table*}
 \caption{List of selected tidal features (+1 Spiral Arm) Found in the WWFI Abell cluster sample. The redshift of the feature was estimated from its host, or if no host redshift was available, the redshift of the cluster.}
 \label{tab:feature-list}
 \centering
 \begin{tabular}{lccccc}
  \hline
    $\alpha_\text{Feature}$ & $\delta_\text{Feature}$ & $z_\text{Feature}$ & $z$ Candidate & Angular Scale & Feature \\
        J2000 & J2000 & & & [kpc/\arcsec] & \\
        (1) & (2) & (3) & (4) & (5) & (6) \\
        \hline
        01:20:54.4 & +19:30:58.6 & 0.0517070 \citep{bilicki+2014} & LEDA 1595431 & 0.661 & Stream \\
        02:56:30.6 & +15:55:22.8 & 0.0352480 \citep{lauer+2014} & UGC 2413 & 0.706 & Stream \\
        10:24:32.5 & +47:50:10.7 & 0.0600400 \citep{alam+2015}  & 2MASX J10243254+4749361 & 1.167 & Stream \\
        10:25:14.3 & +47:44:45.0 & 0.0621100 \citep{ahn+2012} & NVSS J102511+474418     & 1.205 & Stream \\
        11:11:13.8 & +28:44:04.4 & 0.0293400 \citep{ahn+2012} & NVSS J111112+284243     & 0.591 & Tail \\
        12:59:26.6 & +27:59:54.4 & 0.0239070 \citep{rines+2016} & NGC4874                 & 0.485 & Stream \\
        13:00:32.5 & +28:02:01.8 & 0.0212000 \citep{brok+2011} & GMP 2617                & 0.432 & Stream \\
        13:01:04.2 & +27:45:56.0 & 0.0276800 \citep{ahn+2012} & LEDA 83751              & 0.559 & Stream \\
        15:11:55.2 & +05:38:17.4 & 0.0514300 \citep{alam+2015} & 2MASX J15115519+0538171 & 1.010 & Stream \\
        22:49:59.6 & +10:52:23.9 & 0.0787245 \citep{mahdavi+2004} & LEDA 1385552            & 1.491 & Stream \\
        23:12:43.8 & +10:43:21.1 & 0.0221950 \citep{yu+2022} & Mrk 526                 & 0.451 & Tail \\
        23:12:41.7 & +10:44:17.9 & 0.0221950 \citep{yu+2022} & Mrk 526                 & 0.451 & Tail \\
        23:13:29.9 & +10:22:45.9 & 0.0649000 \citep{ledlow+1995} & $z_\text{A2558}$        & 1.255 & Tail \\
        23:13:27.6 & +10:22:27.9 & 0.0649000 \citep{ledlow+1995} & $z_\text{A2558}$        & 1.255 & Tail \\ 
        (23:50:44.6 & +27:16:41.7 & 0.0270370 \citep{yu+2022} & MCG+04-56-014           & 0.546 & Spiral Arm) \\
  \hline
 \end{tabular}
\end{table*}

\subsection{Data Preparation}\label{sec:dataprep}
We start with image cutouts either directly obtained from the LS sky viewer or created from the WWFI stacks. Typical sizes were $3\farcm5\times3\farcm5$. Then we assign the cutout, i.e., the feature, an identification key based on the abbreviation of tidal stream (TS) or tidal tail (TT) and the combination of the central coordinates of the feature (e.g., TS012054+193058 at R.A. $01^\text{h}:20^\text{m}:54^\text{s}$ and decl. $+19\degree:30\arcmin:58\arcsec$). If a bright galaxy overlaps with the feature, an isophotal galaxy model is created (see Section \ref{subsec:subtraction}) and subtracted from the stack. The remaining galaxies and stars are masked and stored in a source mask. Objects that are relatively large compared to the feature and significantly overlap with it are also masked but stored separately in an interpolation mask. In addition, a mask for every source in the image, including the feature, is created, which is used to create the error image (Section \ref{sec:error}).

\begin{figure}
    \centering
    \includegraphics[width=\linewidth]{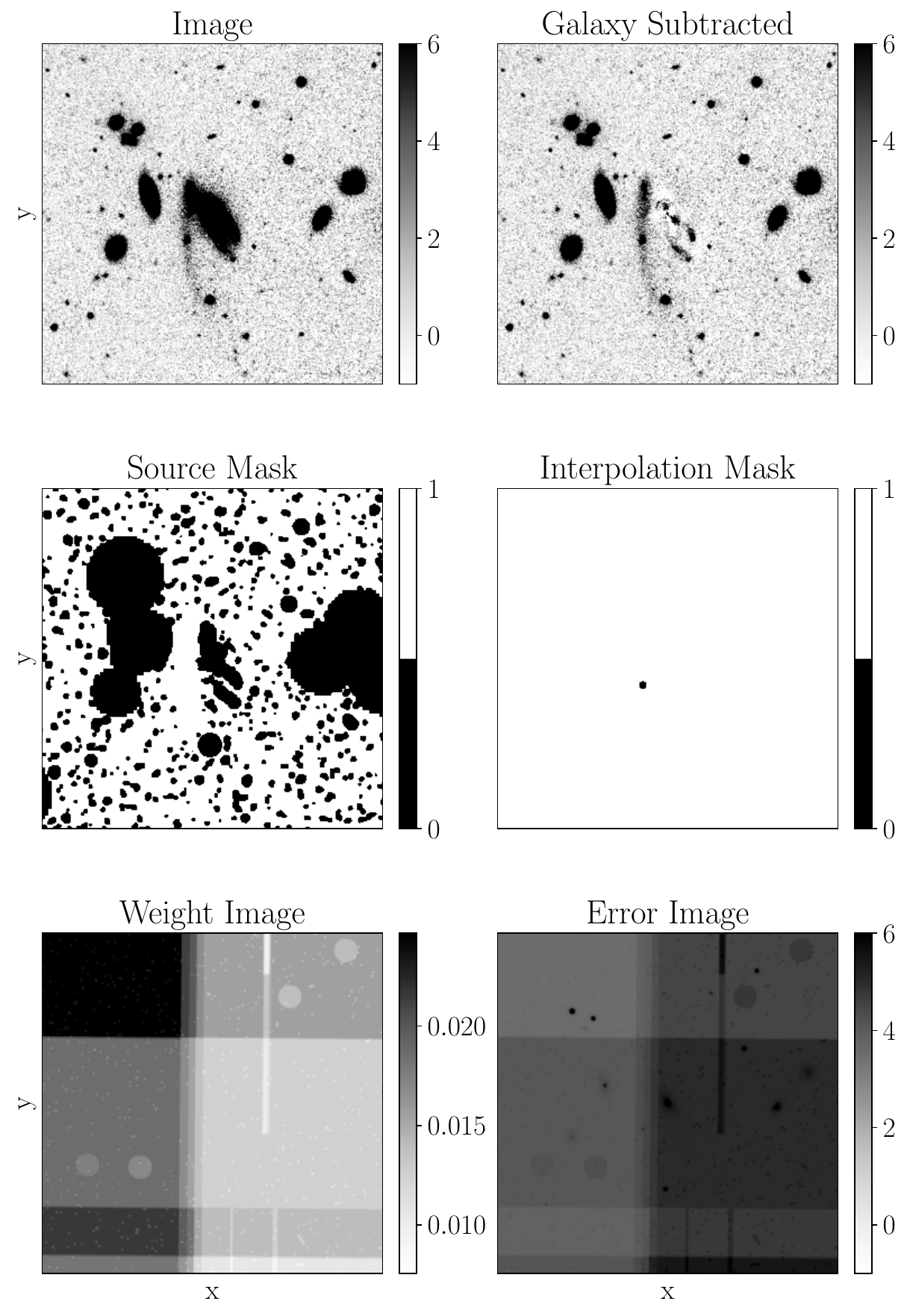}
    \caption{Set of images used for the modeling and photometric measurements. From left to right and top to bottom: (1) The reduced cutout. (2) Same as (1) but after the removal of the galaxy. (3) The source mask. (4) The interpolation mask. (5) The weight image cutout obtained from the data reduction. (6) The corresponding error image, which is determined by Equation \ref{eq:errorimg}.}
    \label{fig:dataprep}
\end{figure}

This set of images (e.g., Figure \ref{fig:dataprep}) for each feature is used for the modeling procedure and photometric measurements.

\section{Methods}\label{sec:methods}
\subsection{Source Masking}\label{subsec:masking}
Masks are created in a semiautomated procedure. First, sources are masked automatically based on the parameters that follow.
\begin{center}
    \textsl{Threshold} $T_0$
\end{center}
The threshold is either a signal-to-noise (S/N) or a surface brightness (SB) threshold. All pixels inside the detection area in the smoothed image with values above the threshold are masked. 
\begin{center}
    \textsl{expand diameter} $\delta$
\end{center}
All individual masks are expanded via a fast Fourier transform convolution with a circular kernel of a given size $\delta$. Every masked pixel is replaced by this kernel, resulting in a more conservative mask.
\begin{center}
    \textsl{border size} $s_\text{border}$
\end{center}
All pixels which are $s_\text{border}$ pixels away from the image border are set to zero, i.e., it performs an image crop. This is useful for wide-field images, where the S/N drops rapidly towards the border. In our case, it was always set to zero.
\begin{center}
    \textsl{detection area} $a$
\end{center}
A circular Gaussian kernel with a given standard deviation $a$ represents the detection area. The background-subtracted image is convolved with it before the sources are detected. A larger parameter results in a higher masked pixel fraction while possibly missing smaller sources.
\begin{center}
    \textsl{background box size} $s_\text{bg}$
\end{center}
With the background box size, the image is divided into grid cells with the size of $s_\text{bg}$. In each cell, the pixel values are $\kappa-\sigma$ clipped with a lower and upper limit of $\kappa=3$ and $\kappa=8$ respectively. After the clipping, the median of the remainder is taken as the background value in that cell. A spline interpolation through the grid is used to create a smoothed background model, which is subtracted from the image. It is also possible to subtract a constant. The constant is determined from all non-zero, non-Nan $\kappa-\sigma$ clipped pixel values, with a lower limit of $\kappa=3$ and an upper limit of $\kappa=4$.

The auto-mask is manually improved by unmasking the feature, expanding masks, and masking missed sources. We should emphasize that it is not necessary to mask every source in the image. Only contaminators near the feature have to be considered. However, for visual demonstration and model quality inspection, it is convenient to have a complete source mask.

\subsection{Galaxy Subtraction}\label{subsec:subtraction}
We subtract isophotal models of galaxies, which either have bright extended halos (large Es or BCGs) or cover a significant fraction of the feature. The models are created with a program that was developed for BCG modeling in \cite{kluge2020} and is based on \textsc{Photutils} \citep{bradley+2023}. The masks are created with the same method as described in Section \ref{subsec:masking}. The ellipse fitting works as follows. First, infinitesimal ellipses are fitted to the inner isophotes, allowing for subpixel precision and averaging along their path. After that, the flux is remeasured from medium to large radii within elliptical rings. To remove statistical effects, i.e., outlier rejection, the median of those rings is taken instead of the mean. Ellipse parameters, such as semi-major axis radius, semi-minor axis radius, central coordinates, ellipticity, and position angle, are varied up to a fixed radius from which only the semiaxis radii are increased. Hence, it is important to note that the ellipses are fitted on changing radii, not on changing surface brightnesses. Both the inner and outer models are then combined into one full galaxy model. 

From strong variations between the boxy and disky isophotes at very small radii, a butterfly/cross-shaped residual can occur. However, this effect is negligible, as only a small percentage of the model is affected, and for most of our sample, the feature does not reside near the center of its host. Still, if a residual arises and overlaps or interferes with the feature, we mask the corresponding region.

In some cases, more than one galaxy needs to be subtracted from the image. Here, we iteratively subtract the galaxies after one another and repeat the process to account for bright overlapping halos that cannot easily be masked. A similar approach is necessary if the feature is still connected to the galaxy. We begin with a first approximate galaxy model and subtract it. Then we improve the mask by having a much better vision of the feature's position than before and create a second model afterward. This strategy improved the quality of the subtracted image significantly.

\subsection{Feature Modeling}\label{subsec:modeling}
As a smaller satellite gets stripped and destroyed by a more massive host, it gets torn apart, leaving a stellar stream. In the absence of triggered star formation, this stream traces the characteristics of the progenitor galaxy, namely brightness, color, stellar mass, age, and metallicity. Even though triggered star formation could modify the stream’s stellar and gaseous properties from the progenitor galaxy, the faint component, i.e., the overall extent of the stream, should remain the same. We define the bright component as concentrated and irregularly shaped regions of brighter flux, e.g., star formation clumps. A similar argument is posed onto tails and its ejected stellar material. Therefore, we use aperture photometry on the features where the aperture shape must be flexible due to the diverse morphologies. Manually drawing the aperture is time-consuming and not reproducible as the visibility of these LSB features strongly depends on how the flux is translated to a color map, e.g., for visual image inspection/analysis. To overcome this issue, we create a model of the feature via an analytic description of the brightness profile. This not only provides exact information about the feature but makes it possible to define a criterion for the shape, i.e., the width of the polygon aperture, in our case the FWHM. We describe our fitting procedure and define our empirically approximated functional form in the following Sections.
\begin{figure*}
    \centering
    \includegraphics[width=\linewidth]{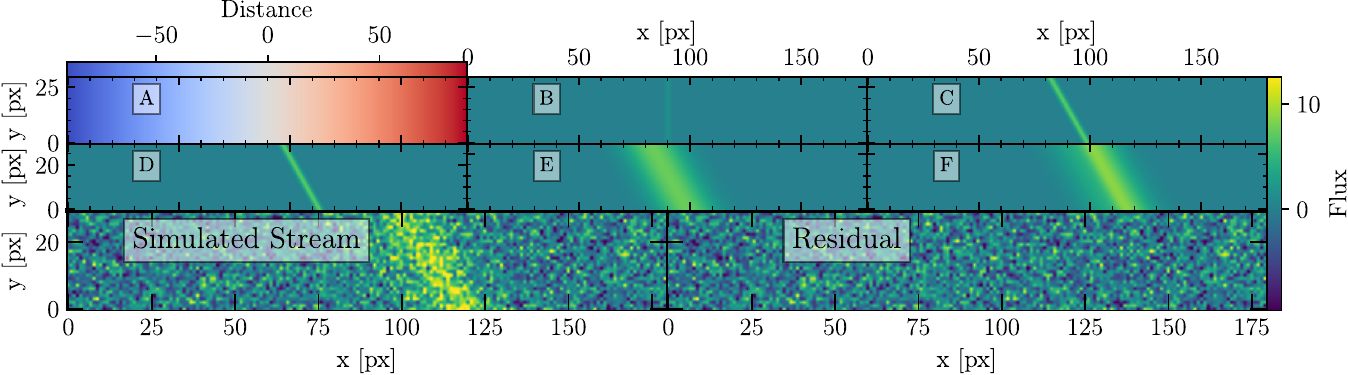}
    \caption{Illustration of the feature modeling. The flux mapping to a color map is identical for all panels except (A). (A) Distance pixel grid (see \ref{app:gaussbox}). A 1 pixel distance corresponds to a value of 1 or -1 respectively. The color map is arbitrary and just for visualization purposes. (B) The distance grid fed into a Gaussian distribution with an amplitude and a standard deviation of 1. (C) The same Gaussian but with a rotated grid and with a higher amplitude. (D) The model position corrected by an $x$-offset. (E) The standard deviation increased. (F) Applied skewness. Note that the orientation and position parameters, i.e., $\alpha$, $x_0$, and $y_0$ are embedded in the grid before the Gaussian is applied.}
    \label{fig:model-method}
\end{figure*}
\subsubsection{Scanner Algorithm}\label{sec:scanner-alg}
The algorithm is based on a simple idea: move a box along the feature and fit an analytic function to every box. The central 1D pixel slice perpendicular to the moving direction then represents the 1D pixel slice of the feature at the central $x$- and $y$-position of the box in the image. The algorithm is part of the author's open source Python package \textsc{astrostreampy}. It is available via PyPi or through GitHub, where a package wrapper is also provided to the user for quick usage. This Section does not explain the package or how to use it but focuses on an in-depth description of the algorithm and its principles. 

First, we must assume that the image is prepared for the modeling as described in Section \ref{sec:dataprep}. Then, an initial box is placed manually on a region of the feature with a high S/N, preferably near the center. From there, the feature is scanned and modeled in both directions. In that way, the box reaches ever lower S/N regimes until the signal drops significantly below the background noise and an analytic function cannot be fitted anymore. Furthermore, the time it takes to finish the model is reduced, as both parts are done in parallel. With that, we maximize the fitted length of the feature. Otherwise, if we start the scanning at one end, most likely a low-S/N region, the fitting procedure could lose robustness and get trapped in a local minimum that badly represents the feature.

Furthermore, plausible box dimensions should be set. For example, if we assume the feature is purely vertical in the image, the width (i.e., the side perpendicular to the feature) should be large enough to cover enough background information. The box height (i.e., the side length parallel) has similar constraints. A small box height could lead to a failure of the fitting as the scatter of the pixel values is too high, especially if the feature is only slightly above the local background. Vice versa, wider boxes increase the S/N, but they also lose information about the structure inside the feature. For a clearer understanding: one can imagine the box being averaged before fitting a Gaussian to it. Hence, a thicker box increases the S/N.

Before we continue with the details of the algorithm, we explain how a model is fitted in each box. The lower-left panel of Figure \ref{fig:model-method} shows a simulated stream segment in a box, which has, in this case, a width of 180 and a height of 30 pixels. The panel next to it is what we want to achieve, a clean residual image with only background noise left. The six smaller upper panels illustrate how the best fit is found. First, a distance grid (A) is created depending on the box dimensions, meaning the pixels in the central vertical slice have a value of zero, and all other pixels have a value regarding their $x$-distance to the center. We encourage the reader to see Appendix \ref{app:gaussbox} for a detailed description of the construction of the distance grid. 

With such a grid, we can produce a feature with a Gaussian profile perpendicular to its direction (B) by inserting the grid into $\exp(-\frac{x^2}{2})$. This Gaussian has an amplitude and standard deviation of 1 and is oriented vertically. Further, we rotate the grid, i.e., the Gaussian by an angle, and increase the amplitude (C). The angle is defined in a way that a positive value rotates counterclockwise. If the box is not centered on the feature, the grid needs to be offset concerning the central $x$- and $y$-position. An $x$-offset is performed, on the grid (D). With an increase of the standard deviation, we widen the stream (E). The last tuning is done by applying a skewness (F) and leaving us with a final stream model. This procedure is repeated for every slice. Of course, if necessary, we can vary the symmetric higher-order moments, i.e., $h_2$ and $h_4$ as well, which would introduce a curtosis.

The parameter fitting for each box is done by the open source Python package \textsc{LMfit-Py}. The \texttt{lmfit.Model} class takes a fit function as an argument and performs a least-squares minimization of the residuals. Apart from just returning to the routine and adjusting the fit parameters, we convolve the model by a Gaussian kernel with the size of the mean point spread function (PSF) across the image, which allows us to fit the intrinsic shape parameters of the stream. While the resulting model image represents the convolved version, the parameter table only contains the intrinsic best-fit values.
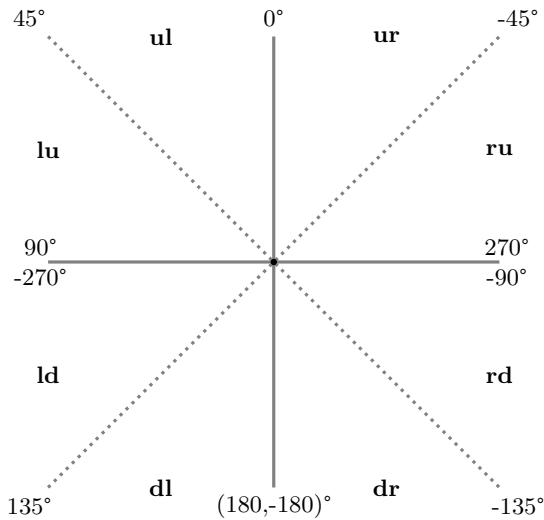
\begin{figure}
    \centering
    \begin{tikzpicture}
        \draw[gray, very thick, dotted] (-3,-3) -- (3,3);
        \draw[gray, very thick,dotted] (-3,3) -- (3,-3);
        \draw[gray, very thick] (0,3) -- (0,-3);
        \draw[gray, very thick] (-3,0) -- (3,0);
        \filldraw[black] (0,0) circle (1pt);
        \node at (-3,-1.5) {\textbf{ld}};
        \node at (3,1.5) {\textbf{ru}};
        \node at (-3,1.5) {\textbf{lu}};
        \node at (3,-1.5) {\textbf{rd}};     
        \node at (-1.5,-3) {\textbf{dl}};
        \node at (-1.5,3) {\textbf{ul}};
        \node at (1.5,-3) {\textbf{dr}};
        \node at (1.5,3) {\textbf{ur}};
        \node at (0,-3.25) {(180,-180)\degree};
        \node at (0,3.25) {0\degree};
        \node at (-3.1,0.2) {90\degree};
        \node at (-3.1,-0.2) {-270\degree};
        \node at (3.1,0.2) {270\degree};
        \node at (3.1,-0.2) {-90\degree};
        \node at (3.25,3.25) {-45\degree};
        \node at (-3.25,3.25) {45\degree};
        \node at (3.25,-3.25) {-135\degree};
        \node at (-3.25,-3.25) {135\degree};
    \end{tikzpicture}
    \caption{Schematic of the sectors, from which the algorithm determines its next box position. The definitions of the bold letters are: \textsl{u}-up, \textsl{d}-down, \textsl{l}-left, and \textsl{r}-right. The angles represent the angle of the stream increasing counterclockwise. The point at the coordinate center resembles the center of the fitting box and each line follows the stream.}
    \label{fig:sectors}
\end{figure}
With the selection of the initial box and its position, an initial model is created. It is used as an "anchor" for both parts of the feature. The code is parallelized, such that both segments are modeled simultaneously using the previous best-fit parameters as initial guesses for the next box. To guarantee that the box remains centered on the stream at every iteration, the peak location of the Gaussian is compared to the center of the box. If there is a discrepancy, the current box center is corrected by either a $x$-, $y$- or $x$-$y$-offset. Based on that, we must tell the box where to move, as it has no global shape information of the feature. By defining a sector map (Fig. \ref{fig:sectors}) with each sector spanning a $45\degree$ angle, we can determine where to move the box based on the fitted grid angle. A keyword is assigned to the sector, e.g., if the feature angle is $78\degree$, the value falls into the "lu" (left up) sector. Of course, “rd” would also be a true sector assignment, which would link to an angle of $258\degree$ or $-102\degree$. The naming is just the combination of the first letters of the primary and the secondary direction and is important for the modeling loop. At each iteration, the two possible sectors are compared to the prior. Based on that, the correct direction is used and translated to the true shift information. This method prohibits random direction flips of the box movement.

In cases where the stream curves beyond $45\degree$ the box dimensions flip, such that the shorter side remains parallel to the feature. Additionally, the width and height of the box change smoothly with the measured stream angle. This allows for a continuous transition from a vertical to a horizontal box, or vice versa. At the limit of $45\degree$, the box becomes a square. The widths and heights are determined by two inverse tangent functions and are limited by the dimensions of the initial box. It needs to be noted that this approach slowly fails if the stream angle in the initial box deviates strongly from a purely horizontal or vertical feature segment. In the case of the sample used in this work, it was no issue, as the features remained almost straight. Hence, the use of dynamic box dimensions was turned off during the model creation.

As soon as the best-fit values are determined, a box model, i.e., the model of the current segment is created. The central 1D pixel slice of that model is then used as a model slice of the whole feature at that $x$- and $y$-coordinate. The slice used is always along the larger box side. With the step size being 1 pixel no gaps or overlaps are created, without considering a dimension flip. To account for those cases, we take the average of the current model and the temporary model carrying the current 1D slice. Everything described up until now is repeated in each step of the modeling. 

The \textsc{LMfit-Py} library model fitting is designed to continue with previous parameters if no possible solution is found. To counter that behavior, we introduce termination thresholds:
\begin{itemize}
    \item \textit{Non-Gaussian Best Fit}: If consecutively (three times in a row) no peak in the model, i.e., no Gaussian fit, was found, the modeling terminates.
    \item \textit{User Fixed Endpoints}: If the current box center is near ($n$ pixel) one of the two endpoints, the modeling terminates after continuing for $2n$ steps.
    \item \textit{S/N}: If the ratio of the fitted amplitude of the Gaussian and its fitting error is below a certain value, the modeling terminates.
    \item \textit{Repeating parameters}: If the fitted amplitude of three successively fitted amplitudes is within a certain threshold, the modeling terminates.
\end{itemize}

If the endpoints are given, only the non-Gaussian threshold is checked; otherwise, we brute-force the modeling. This might introduce badly modeled end segments, but in general, those can be cut off later while inspecting the model quality and using the \textsc{astrostreampy} built-in model modifier.

It should be noted that our approach with fixed boxes (i.e., their sides are always parallel to the image borders) might not be the best solution to the problem. It rather might be better to rotate the box and always keep it perpendicular to the feature. That would be a big improvement, as the dimensions stay the same, but additionally makes the problem more complex, as we cannot easily just take the central slice as a model anymore. While our approach is sufficient for features with a small curvature, the exact opposite poses a limit to our method. This is a limitation that might get updated in the future to also work for strong curved features.

\subsubsection{Gaussian Fitting}\label{subsubsec:gaussianfit}
In a first-order approximation, we use a Gaussian as the analytic fit function for every 1D slice of the feature. There might be a better solution, but we will show that our approach yields very promising results. However, for some features, a standard Gaussian is not sufficient, and they show substructure and altered shapes, possibly caused by star formation and gravitational forces acting upon the feature. 

To account for these shape deviations, i.e., skewness (tilt) and curtosis (”taildness”) or combinations of both, we use the Hermite basis functions. These functions are derived from a set of orthogonal polynomials, called the Hermite polynomials, and are defined as shown below.
\begin{equation}\label{eq:chap-modeling:hermitepols}
    H_n\left(x\right) = \left(-1\right)^n e^{x^2} \frac{d^n}{dx^n} e^{-x^2}
\end{equation}
The order of the polynomial is given by $n \in\mathbb{N}$. Normalizing them by $1/\sqrt{2^n n!}$ and multiplying it with an amplitude parameter ($h_n$) we arrive at the Hermite basis functions. 
\begin{align}\label{eq:chap-modeling:hermitbasis}
    H_2\left(x\right) &= \frac{h_2}{\sqrt{8}} \left(4x^2-2\right) = \frac{h_2}{\sqrt{2}} \left(2x^2-1\right) \\ 
    H_3\left(x\right) &= \frac{h_3}{\sqrt{48}} \left(8x^3-12x\right) = \frac{h_3}{\sqrt{3}} \left(2x^3-3x\right)\\ 
    H_4\left(x\right) &= \frac{h_4}{\sqrt{384}} \left(16x^4-48x^2+12\right) \\
    &= \frac{h_4}{\sqrt{24}} \left(4x^4-12x^2+3\right)
\end{align}
In spite of that, the basis functions can create negative values in the model if the $h_n$ deviates strongly from zero ($\gtrsim\pm0.1$). To reduce this issue, we use a cumulative distribution function (CDF) instead of $H_3$. We use the CDF from \textsc{SciPy} (\texttt{scipy.stats.norm.cdf}), which is defined as 
\begin{equation}
    \text{CDF}(x,\sigma,s) = s\int_{-\infty}^{x} \frac{1}{\sqrt{2 \pi \sigma^2}} \, e^{ -\frac{(t - \mu)^2}{2 \sigma^2} } \, dt\,,
\end{equation}
where we add an additional amplitude parameter $s$, which should be equivalent to the $h_n$. The argument to use the CDF instead of the third Hermite basis function is that $H_3$ is most prone to fall below zero. Generally, negative values are allowed, as the local background should scatter around zero, but strong deviations are nonphysical for the light profile. Our final fit function is as follows:
\begin{equation}\label{eq:fitfunc}
    F=A\Bigl[e^{-\frac{1x^2}{2\sigma^2}}(1+\text{CDF}(x,\sigma,s)+\sum_{i\in\{2,4\}}H_i(x/\sigma,h_i)\Bigr] + b
\end{equation}
We summarize all fit parameters of the modeling in Table \ref{tab:allparams}.
\begin{table*}
    \centering
    \caption{List of the Fit Parameters of the Gaussian Modeling. The first three rows are the fit parameters of the grid (see Appendix \ref{app:gaussbox}). The rest are those of the analytic fit function (Equation \ref{eq:fitfunc}).}
    \label{tab:allparams}
    \begin{tabular}{cc}
        \hline
        Parameter & Description \\
        \hline
        $\alpha$ & Position angle of the feature inside the fitting box. \\
        $x_0$ & The $x$-offset of the center of the feature inside the fitting box. \\
        $y_0$ & The $y$-offset of the center of the feature inside the fitting box. \\
        \hline
        $A$ & The global amplitude of the Gaussian.\\
        $\sigma$ & The standard deviation of the Gaussian. \\
        $b$ & The background offset of the Gaussian. \\
        $h_2$ & The amplitude of the second-order Hermite basis function ($H_2$). \\
        $s$ & The amplitude of the CDF resembling the third-order Hermite basis function ($H_3$). \\
        $h_4$ & The amplitude of the fourth-order Hermite basis function ($H_4$). \\
        \hline
    \end{tabular}
\end{table*}

\subsection{Photometric Measurements}
To measure fluxes of tidal features, a good aperture has to be chosen. However, the complex shape does not allow for a simple method, such as one circular aperture governed by a radius. Drawing a polygon by hand is one possibility, but leaves room for a large variation in shape. Not only does human accuracy vary, but it is also difficult to decide whether the polygon is too wide, too narrow, too long, or too short. In addition, manual processes are time-consuming, especially clicking on points to accurately follow a shape. A more robust and automated method is therefore required, especially to enable comparisons between same objects but different studies; a method that ensures that all feature pixels are within the aperture. 

We use the best-fit parameters of each Gaussian (Equation \ref{eq:fitfunc}) to calculate the FWHM for each slice along the feature. From now on, we use $\Gamma_n$ as a symbol for the FWHM where $n$ represents the multiples of the FWHM; e.g., $\Gamma_3$ would mean the flux inside three FWHM. From there, slice by slice, all pixels farther than half the $\Gamma_n$ from the peak are set to zero. The other pixels are set to one. This is enough to create an aperture mask. Of course, the masking threshold can be adjusted to create wider or narrower apertures. In addition, we smooth the aperture by using a running mean along the $\Gamma_n$ of all slices. The apertures are used to measure the key properties of the features, i.e., apparent and absolute magnitude, surface brightness, and width. We use $\Gamma_1$ for the central surface brightness and color, and $\Gamma_3$ for apparent and absolute magnitude.

Our goal is to create an aperture mask that encloses almost all of the feature’s flux. Therefore, we establish a 76–98–99.9 rule. For a standard Gaussian, within $\Gamma_3$, $\sim99.96\%$ of the distribution is covered (Appendix \ref{app:rule}). Even if the true percentage for higher-order Gaussians differs, it only adds values from the wings, which should scatter anyway around the mean local background, ideally zero. As such, we defined a robust way for feature apertures from Gaussian models.

There are cases in which galaxies or stars overlap with the feature. If the feature is broad enough and the sources are relatively small, those can be masked, and \textsc{astrostreampy} will still be able to perform a good fit. Other scenarios involve either too many or too large sources and a relatively narrow feature. Both imply that too much area would be masked, and the fitting would fail in those regions. Those sources, which strongly overlap, are then interpolated with a 2D Gaussian Kernel and a position angle ensuring that these pixels are interpolated along the stream. The position angle is important such that the interpolation happens along the aperture and allows for asymmetric kernels with different standard deviations for $x$ and $y$. For photometric measurements, all pixels that were either used as a source mask or an interpolation mask are then filled with the corresponding information from the created model or remain zero. We rely on this approach, as measurements on the model and the true data differ on a $\sim5\%$ level.

We measure the surface brightness as well as the apparent and absolute magnitudes for each tidal feature within an aperture that is based on the FWHM criterion. The flux inside the aperture is calculated via
\begin{equation}
    F = \sum_{x=0}^n\sum_{y=0}^n(I_{xy}-\Bar{\beta})A_{xy}
\end{equation}
where $I_{xy}$ is the data pixel value, $\Bar{\beta}$ the local background value, $A_{xy}$ the pixel value of the aperture mask (image) with $A_{xy}\in\{0,1\}$. Further,
\begin{align}
    &m = -2.5\log(F) + \text{ZP}\\
    &SB = -2.5\log\Biggl(\frac{F}{n\times\theta^2}\Biggr) + \text{ZP}
\end{align}
holds, where $\theta$ is the pixel scale of the image in \arcsec pixel$^{-1}$ and $n$ is the number of nonzero pixels inside the aperture given by $\sum_{x=0}^n\sum_{y=0}^n A_{xy}$. To obtain absolute magnitudes, we subtract the distance modulus $M = m-\mu$. Surface brightnesses are corrected for cosmological dimming via $F=(1+z)^4 F$.

The local background is directly determined from the best-fit parameters. For each slice of the feature, we already have the information of the background level. We compute a weighted average of all offsets to obtain the mean local background $\Bar{\beta}$. The quality of $\Bar{\beta}$ then strongly depends on the number of available background pixels for each box. To some extent, the uncertainty of this approach and the resulting background value are captured with a Monte Carlo Error Propagation (MCEP). We also encapsulate the mean background uncertainty in our flux error estimation using the standard error of the mean of all fitted offsets.

The effective surface brightness is measured at the effective radius. For that, we create another aperture, based on the model, where each pixel that is not at a distance of the effective radius from the center of feature, is set to zero. With this approach we get a 1 pixel wide border aperture, at which the surface brightness is measured.

\subsection{Color Measurements}
The BCG sample of \cite{kluge+2020} only covers the $g'$ band. However, for some clusters, follow-up observations were performed for other studies, e.g., ultra-diffuse galaxy (UDG) populations in A262 and A1656 (Coma Cluster) in \cite{zoeller+2024}. The latter is for now the only cluster where the $r'$ band of the sample is available. For all others, we utilize Legacy Survey DR10 data \citep[LS-DR10][]{dey+2019}. Eleven out of the 12 remaining features were covered by the LS-DR10 footprint. We extract cutouts and repeat the data preparation process for the LS $r$ and $g$ band. Further, feature models and photometric apertures are created with \textsc{astrostreampy}. We compare each image, model, residual, and aperture in every filter to decide which aperture is best for the color measurements. The strongest criteria here are the quality of the aperture mask and how well it traces the shape of the stream and does not show any artifacts from the creation process, i.e., missing parts or unwanted shapes.

Before we apply the final selected aperture, we discard every pixel that was either masked or interpolated (see Section \ref{sec:dataprep}) in every band. In that way, we ensure that first, only the same pixels are measured, and that we only measure the true color of the feature. Similar to color measurements of stars, we do not use the complete feature, as we assume that the color measurement is most reliable in the inner regions, i.e., inside one FHWM. Further for the WWFI colors, we use the ZP$_{50}$, instead of ZP$_\text{inf}$, and the standard ZP of the Legacy Survey of 22.5 mag.

The colors are used to transform measured $g'$ brightness into the $V$ band following \cite{jester+2005}, which is necessary for comparisons with various galaxy types in the $M_\mathrm{tot} - r_e$, $\mu_e - r_e$ and $M_\mathrm{tot}-\mu_e$ parameter spaces. We have to compensate for discrepancies between the WWFI $g'$ (Pan-STARRS) and the Sloan Digital Sky Survey (SDSS) $g$ band via $g=g'+0.08$ and $r=r'+0.01$ for the Sun \citep[see][]{tonry+2012,willmer2018}. This results in Equations \ref{eq:vwwfi} and \ref{eq:vls}, which transforms $g'$ into $V$ depending on the used color. The total absolute $V$ band magnitudes are further utilized to calculate the luminosity of the features in units of solar luminosities via $L=10^{0.4(M_{V,\odot}-M_V)}L_\odot$, where $M_{V,\odot}=4.80$ is the solar absolute $V$ band magnitude from \citep{willmer2018}.
\begin{equation}\label{eq:vwwfi}
    V_\mathrm{WWFI}=g'-0.59(g'-r')+0.0287
\end{equation}
\begin{equation}\label{eq:vls}
    V_\mathrm{LS}=g'-0.59(g-r)+0.07
\end{equation}

Additionally, for comparisons with cluster members in the Coma cluster, we apply a correction of the distance modulus and subtract a $K$ correction term from the total absolute magnitude. In this manner, the measured brightnesses are converted into one rest frame. All $K$ terms were determined with a calculator \citep{chilingarian+2012}.

\subsection{Error Evaluation}\label{sec:error}
\subsubsection{Modeling Uncertainties}
Final brightnesses mainly depend on the quality of the model and its $\Gamma_n$ aperture. Hence, we estimate additional uncertainties introduced by the modeling algorithm. Covariance matrices are directly provided by the used least-square-minimization package and utilized to perform an MCEP to quantify the impact of the best-fit parameters on the size and shape of the aperture mask. One stream model is created to obtain a sample of best-fit parameters for each slice. Then, 25 times for each slice of the feature, a new best-fit parameter set (bottom half of Table \ref{tab:allparams}) is drawn from a multivariate distribution governed by the covariance matrix. From there, we have 25 slightly different apertures, measure the magnitudes, and take the standard deviation of all results as an error representing an aperture error. This procedure was performed for two different streams from Table \ref{tab:feature-list}, one simple (TS012054+193028) and one complex (TS125926+275954), with higher-order terms enabled. For both, the 25 measurements for slightly different apertures scatter within a standard deviation of 0.005 magnitudes.

Not only do the fit parameters themselves hold an uncertainty, but the modeling depends on where the initial box placed and what its dimensions are. This robustness was captured by creating 50 models of the same feature with different initial boxes. The first box was set manually. Then, its width, height, and position were randomly varied within a certain range. For every model, an aperture was created, and the corresponding apparent magnitude and surface brightness were measured. The standard deviation of these measurements represents the robustness of the algorithm (Section \ref{sec:robtest}). For the total brightness, we find a scatter of $\Delta m_\sigma = 0.009$ mag and for the surface brightness, a value of $\Delta SB_\sigma = 0.013$ mag arcsec$^{-2}$.

\subsubsection{Photometric Uncertainties}
Flux errors inside the apertures are calculated from error images that contain the flux uncertainty for every pixel. We follow a method to generate error stacks from the stacked science images, which is described in detail in \cite{zoeller+2024}. The flux uncertainty in a pixel at positions $x$ and $y$ can be approximated by
\begin{multline}\label{eq:errorimg}
    E(x,y) = \Biggl\{ \frac{|\mathcal{D}(x,y)|\times\text{median}\{\mathcal{W}\}}{\mathcal{G}\times \mathcal{W}(x,y)} + \\\Biggl(\frac{\text{std}\{\mathcal{D}_m \times \sqrt{\mathcal{W}_m/\text{max}\{\mathcal{W}\}}\}}{\sqrt{\mathcal{W}(x,y)/\text{max}\{\mathcal{W}\}}}\Biggr)^2 \Biggr\}^{1/2}\,,
\end{multline}
where $\mathcal{D}$ is the image, $\mathcal{G}$ is the gain of the camera, $\mathcal{W}$ is the weight image, $\mathcal{D}_m$ is the masked image, and $\mathcal{W}_m$ is the masked weight image. For the WWFI stacks, weight images are a byproduct of the data reduction and store information on the spatially varying depth and, consequently, spatially varying S/N. For the Legacy Survey, we assume a constant weight over the entire FOV. The error flux inside the feature apertures is measured and added/subtracted to the feature's flux and translated into brightness errors.
\begin{align}
    &\Delta F = \sum_{i=0}^n\sum_{j=0}^n E_{ij}A_{ij} \\
    &\Delta SB^\pm = \Delta m^\pm = \Delta M^\pm = -2.5 \log \Biggl(\frac{F \mp \Delta F \mp \Delta \beta}{F}\Biggr)
\end{align}
Here, $\Delta \beta$ is the cumulative background uncertainty of the local background estimated from the standard error of the mean of all Gaussian offset errors of every model slice, multiplied by the number of pixels inside the aperture. Note that the uncertainties for the surface brightness and total magnitudes still differ because the surface brightness flux is dimmed by the mentioned factor.

We overestimate the global brightness errors by adding the modeling uncertainties to the photometric error. This error is propagated to the color error via
\begin{equation}\label{eq:deltacolor}
    \Delta(g'-r') = \sqrt{\Delta g'^2 + \Delta r'^2}\,.
\end{equation}

The uncertainty of the width of the stream, both the $\Gamma_1$ width $w$ and the effective radius $r_\mathrm{eff}$, depend on the integer-based construction of the corresponding apertures and from which the widths are calculated, and the scatter of the mean widths of the apertures created via the MCEP. Both add up to a width error of $\Delta w = 2$ pixels, where we assumed the MCEP error to be consistent for all features.

\section{Modeling Performance}
\subsection{Model Quality}
\begin{figure*}
    \centering
    \includegraphics[width=\linewidth]{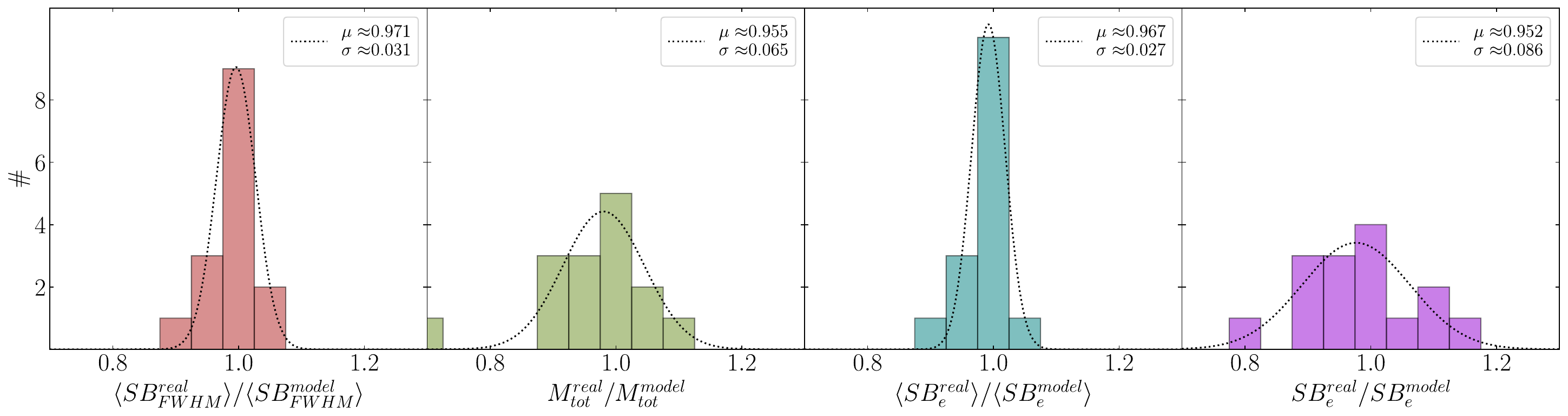}
    \caption{Histograms of the flux ratios of the feature measurements on the real data and the created model. The dotted line represents the fitted normal distribution to each measurement with the corresponding mean $\mu$ and standard deviation $\sigma$ displayed in the legend. Note that we subtract the fluxes and not divide magnitudes. The $x$-axis label is a convenience for the reader.}
    \label{fig:precisions}
\end{figure*}

\begin{figure}
    \centering
    \includegraphics[width=\linewidth]{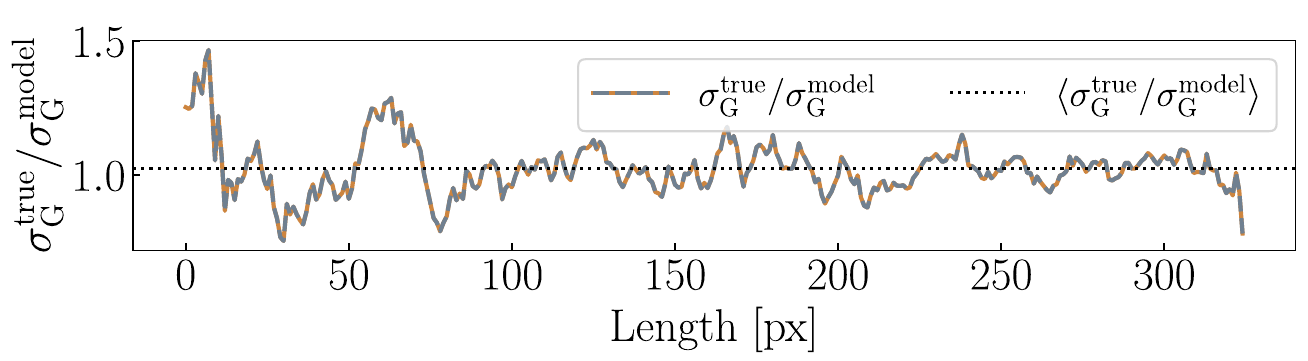}
    \caption{Ratio of the best-fit standard deviations of the Gaussian slices along TS012054+193028 of the real feature and the injected model. The mean ratio of $\sim1.02$ is given by the dotted line.}
    \label{fig:widthprecision}
\end{figure}

In addition to visual confirmation of a successful model from either the residual image or a slice inspection, we quantify the fit quality by the ratio of the modeled and measured flux. If we compute the ratio of the measured fluxes on the real image and the created model within the same aperture, we are able to extract how precise we get in our (small) sample. Figure \ref{fig:precisions} shows four histograms, one for the surface brightness $SB$, the total absolute magnitude $M_\mathrm{tot}$, the surface brightness inside the effective radius $\langle SB_\mathrm{e} \rangle$, and the surface brightness at the effective radius $SB_\mathrm{e}$. The mean surface brightness inside an aperture is less prone to outliers, i.e., noise peaks or other effects. We reach an accuracy of $\approx 3\%$ in $\langle SB_\mathrm{e} \rangle$. Otherwise, $M_\mathrm{tot}$ and $SB_\mathrm{e}$ are affected strongly, especially the latter, which is an annulus with a width of just 1 pixel. Hence, the accuracy is smaller with $\sigma \approx 9\%$. Analogous to that, we constraint the model quality with respect to the width of the feature. We utilize the best-fit parameters of the model (true) of TS012054+193028 and its injection (model), which was the model with added noise and remodeled by \texttt{astrostreampy} and investigate the Gaussian standard deviation $\sigma_G$ fit parameter (see Figure \ref{fig:widthprecision}). The ratio of the true and model width profile, i.e., $\sigma_\text{G}$ as a function of the slice position, is computed, and its average represents the mean precision of the mean stream width along its full extent, which is $\overline{\sigma_\text{G}} = \sigma_\text{G}^\text{true}/\sigma_\text{G}^\text{model}\approx3\%$.

\subsection{Influence of the Initial Box}\label{sec:robtest}
\begin{figure}
    \centering
    \includegraphics[width=\linewidth]{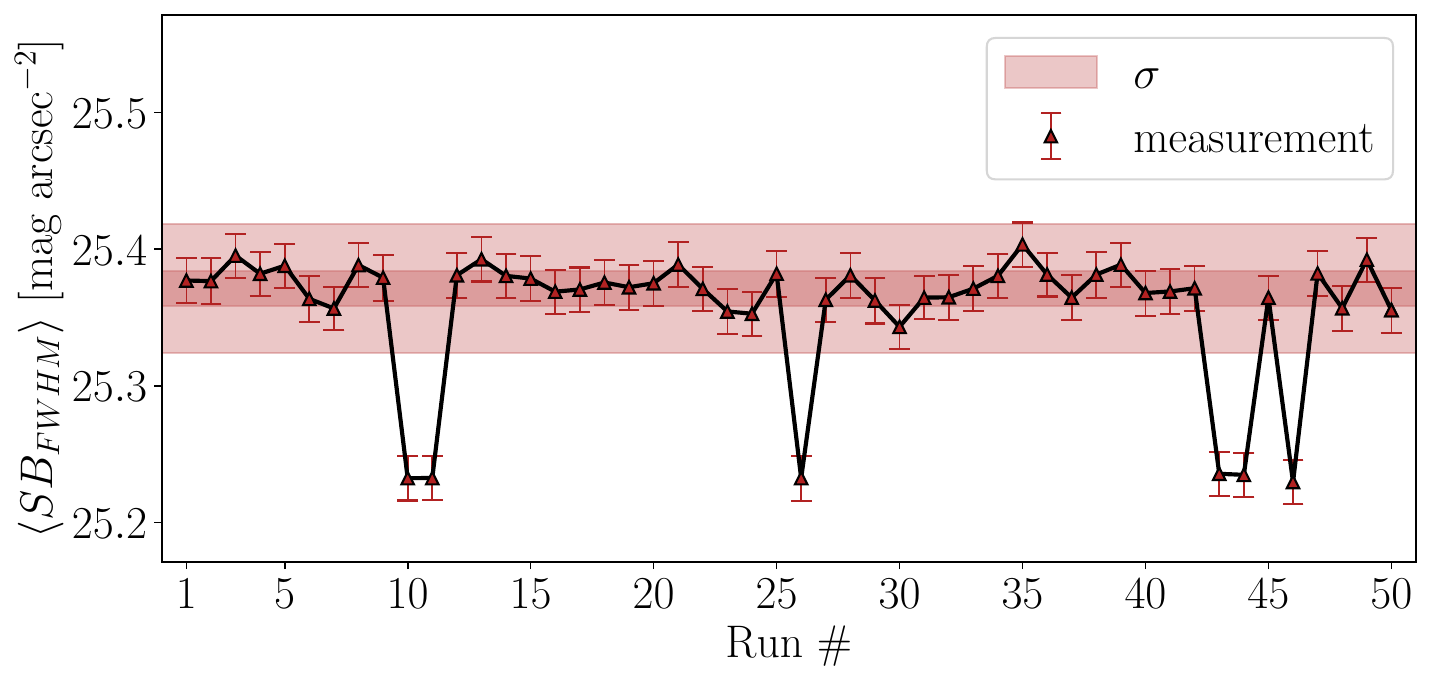}
    \caption{Surface brightness robustness of 50 TS012054+193028 stream models. Measurements are shown in black triangles with corresponding brightness error bars in red. Two $1\sigma$ error stripes are overlaid, of which the more opaque one is the standard deviation of all data points and the less opaque without the outliers. Note that the error bars of the surface brightnesses are only coming from the flux error image and are not the same errors computed for the final results.}
    \label{fig:robustness}
\end{figure}

We create 50 models of the stream TS012054+193028 with different initial boxes. The central coordinates, width, and height are changed randomly by a few pixels. We set the maximum number of model iterations to 200 to ensure that the full extent of the stream is captured,  without reaching regions with low S/N that are prone to result in erroneous models. For each of the 50 models, we create $\Gamma_1$ apertures and measure the surface brightness and compute the corresponding flux uncertainty. We show the result in Figure \ref{fig:robustness}. Additionally, we compute the median ($\rm SB\approx25.3713$ mag arcsec$^{-2}$) and standard deviation ($\Delta \rm SB \approx0.0474$ mag arcsec$^{-2}$) of the 50 measurements and plot it as a $1\sigma$ horizontal error stripe. Six outliers show brighter values, which are caused by earlier modeling terminations of the lower/fainter part of the stream. Hence, the surface brightness inside the aperture is brighter. We emphasize that this analysis is performed without checking the models, making it a blind check of the robustness. If we discard these six points, the standard deviation reduces to $\sigma \approx 0.013$ mag arcsec$^{-2}$. We see that the scatter of the surface brightnesses is comparable to the measured flux uncertainty. The same can be applied to the total brightness, which results in a clipped scatter of $\sim0.009$ mag. Both values are added to cumulative brightness errors.

\subsection{Injection Recovery Test}
\begin{figure}
    \centering
    \includegraphics[width=\linewidth]{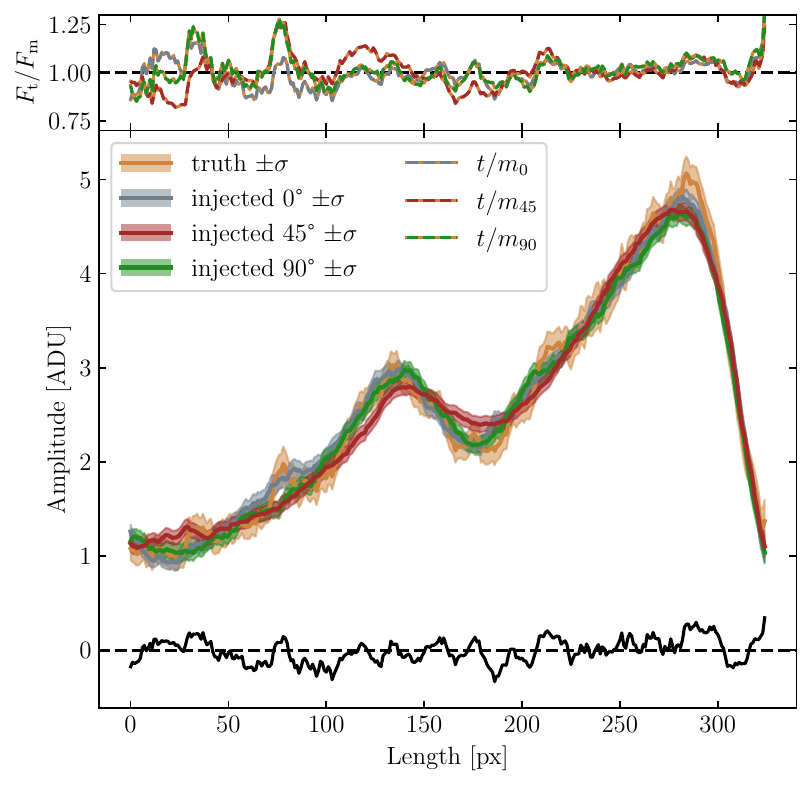}
    \caption{Fitted amplitude parameters of four TS012054+193028 models. Brown represents the model obtained from the original image. Gray is the same model injected into Gaussian noise. The same was done for green and red, but the model was rotated by 90 and $45\degree$ respectively. The residual from the model and the injected $0\degree$ model is shown in black. The mean of this residual is displayed by the black dashed line. The ratios of the flux profiles of the $0\degree$ to the $45\degree$ and $90\degree$ model are shown in the top panel. Here, the black dashed line shows a constant ratio of $1$.}
    \label{fig:injection}
\end{figure}

Gaussian noise is introduced to one of our feature models (truth) and used to remodel it with the algorithm. We compare the best-fit parameters and fit uncertainties of both models. A comprehensive analysis for the amplitude parameter $A$ is presented in Figure \ref{fig:injection}. We show the truth in orange and three injections with different orientation angles, i.e., $\alpha\in\{0\degree,45\degree,90\degree\}$. In all scenarios, we manage to retrieve the best-fit parameters for the amplitude profile within a $2\sigma$ interval from the original model. The $x$-axis is denoted as "length", and corresponds to the pixel information on the image, specifically, the centers of the boxes along the feature. In this context, the $y$ box centers align with the vertically oriented stream, while all other orientations are mapped onto the same coordinate system.

\section{Results}\label{sec:results}
\subsection{Stream and Tail Catalog}
One objective of this study is to compile a catalog detailing the structural characteristics of stellar tidal streams. While a listing of stream features around nearby galaxies exists in the SSLS \citep{delgado+2023b} and in \cite{carretero+2023}, those only encompass surface brightnesses and colors. We build a catalog of different streams up to $z\sim0.08$ and incorporate additional physical properties, which are not present in the mentioned literature catalogs. Additionally, in this work we add tails for a more comprehensive overview. 

Although the number of features examined in this study is manageable, the total number of detected features is anticipated to rise significantly with forthcoming deep surveys due to their increased depth. Therefore, establishing a standardized naming convention is imperative to ensure clarity and distinction. One proposed naming convention involves employing a general abbreviation followed by the R.A. and decl. coordinates. For stellar tidal streams, we advocate the format outlined below, as already adapted throughout this work.

We provide an example for the stream TS012054+193028. Here, '012054' represents the R.A. of 1$^\text{h}$, 20$^\text{m}$, and 54$^\text{s}$, while '+193028' signifies the decl. of $+19\degree$, $30\arcmin$, and $28\arcsec$. The feature coordinates are chosen to be the center of it, i.e., the point halfway along the feature. Note that for fully/multiple looped streams, this is not straightforward anymore. Here, one could choose either any point on the feature or its centroid. The prefixes 'TS' and 'TT' denote ``Tidal Stream'' and ``Tidal Tidal'', respectively.

Following the steps described in Section \ref{sec:methods}, we present the feature properties in our WWFI stream and tail catalog in Table \ref{tab:stream-props}. Direct comparison to the stream sample in the SSLS \citep{delgado+2023b} is not possible because there is no overlap with the WWFI cluster survey \citep{kluge2020}. However, we can generally compare the measured $g'$-band surface brightnesses and $(g'-r')$ colors. We measure mean stream surface brightnesses of $\sim 26.25$ $g'$ mag arcsec$^{-2}$, which is brighter than the average $\sim 26.9$ $g$ mag arcsec$^{-2}$ from the SSLS. This is most likely caused by our different choice of where we measure the surface brightness. We average the surface flux in the central $\Gamma_1$ region, whereas circular apertures along the stream are used in the SSLS, which includes the full area. 

In addition, we compute mean detection depths following \cite{roman+2020}:
\begin{equation}\label{eq:depths}
    \epsilon = -2.5 \log \Biggl(3\frac{\mathrm{std}\{\mathcal{D}_\mathrm{m}\}}{7\arcsec\times\theta}\Biggr) + \mathrm{ZP}
\end{equation}
where $\mathcal{D}_\mathrm{m}$ is the masked image, $\theta$ the pixel size of the image in arcseconds and the $7\arcsec$ comes from the typical size of the features in this sample. It computes the surface brightness that results for a flux level of three times the standard deviation of the background for the given area. The SSLS is deeper $\epsilon_\mathrm{SSLS}=28.89$ mag arcsec$^{-2}$ than our sample $\epsilon_\mathrm{WWFI}=28.38$ mag arcsec$^{-2}$. Hence, we attribute the larger fluxes in our work to selection effects. Regarding $(g'-r')$ colors, our results are consistent with the streams in the SSLS, namely all share similar values, ranging from $0.49-0.96$ (SSLS) and $0.45-0.90$ (this work).

We note that these measurements are also slightly affected by the choice of the redshift of the feature. As we do not have photometric or spectroscopic redshifts of the features directly, we assumed either the cluster redshift or that of the most probable host galaxy.
\begin{table*}
  \centering
  \caption{List of 14 tidal features (+1 Spiral Arm) and Their properties detected in the 170 Wide Field Abell cluster observations with the Wendelstein Telescope. Column (1) gives the suggested general name of the feature. Column (2) shows the $g'$ band surface brightnesses inside $\Gamma_1$. Columns (3) and (4) show absolute $g'$ and $V$-band magnitudes. (5) $(g'-r')$ color. Column (6) ´shows Solar luminosities. Columns (7) and (8) show the effective $g'$-band surface brightness at the effective radius. Columns (9-10) show the shape properties' length and width.}
  \label{tab:stream-props}      
        \setlength{\tabcolsep}{2pt}
        \scriptsize
        \begin{tabular}{c c c c c c c c c c c c c c c}
            \hline
            \hline
            Stream & SB$_\text{FWHM}$ & $M_\mathrm{tot}$  & $M_\mathrm{tot}$ & $(g'-r')$ & $L$ & $\mu_\text{eff}$ & $r_\text{eff}$ & $w$ & $l$ \\
            & $(g'$ mag arcsec$^{-2})$ & $(g'$ mag) & (V mag) & (mag) & ($10^9 L_\odot$) & (V mag arcsec$^{-2}$) & (kpc) & (kpc) & (kpc) \\
            (1) & (2) & (3) & (4) & (5) & (6) & (7) & (8) & (9) & (10) \\
            \hline
            \\
            TS102432+475010 & $26.949\pm0.026$ & $-14.97\pm0.04$ & $-15.2\substack{+1.0\\-0.5}$ & $0.5\substack{+1.6\\-0.8}$ & $0.1\pm0.06$ & $27.017\pm0.026$ & $1.2\pm0.5$ & $3.0\pm0.5$ & $33.5\pm1.7$ \\
            TS102514+474445 & $26.632\substack{+0.026\\-0.025}$ & $-16.88\pm0.04$ & $-17.4\substack{+0.4\\-0.3}$ & $0.9\substack{+0.7\\-0.5}$ & $0.74\substack{+0.22\\-0.21}$ & $26.578\pm0.025$ & $2.9\pm0.5$ & $8.4\pm0.5$ & $49.8\pm1.7$ \\
            TT111113+284404 & $24.848\pm0.019$ & $-18.075\pm0.016$ & $-18.31\substack{+0.08\\-0.07}$ & $0.52\pm0.12$ & $1.75\pm0.12$ & $25.056\pm0.019$ & $1.97\pm0.24$ & $4.38\pm0.24$ & $44.8\pm0.9$ \\
            TS125926+275954 & $26.374\pm0.023$ & $-18.486\substack{+0.026\\-0.025}$ & $-18.85\pm0.04$ & $0.67\pm0.04$ & $2.89\pm0.09$ & $26.452\pm0.023$ & $5.06\pm0.2$ & $15.91\pm0.2$ & $95.9\pm0.7$ \\
            TS130032+280201 & $27.049\pm0.023$ & $-14.872\substack{+0.028\\-0.027}$ & $-15.32\pm0.04$ & $0.8\pm0.05$ & $0.111\pm0.004$ & $27.061\pm0.023$ & $1.83\pm0.18$ & $5.44\pm0.18$ & $24.5\pm0.7$ \\
            TS130104+274556 & $26.601\pm0.021$ & $-16.872\pm0.021$ & $-17.14\pm0.04$ & $0.5\pm0.06$ & $0.596\substack{+0.021\\-0.022}$ & $26.648\pm0.021$ & $4.17\pm0.23$ & $12.41\pm0.23$ & $37.8\pm0.8$ \\
            TS012054+193028 & $25.256\pm0.022$ & $-18.32\pm0.024$ & $-18.72\substack{+0.12\\-0.11}$ & $0.79\substack{+0.2\\-0.18}$ & $2.55\substack{+0.26\\-0.27}$ & $25.189\pm0.022$ & $2.2\pm0.5$ & $6.7\pm0.5$ & $69.4\pm1.5$ \\
            TS151155+053817 & $26.469\pm0.021$ & $-17.453\pm0.021$ & $-17.88\substack{+0.15\\-0.13}$ & $0.84\substack{+0.24\\-0.21}$ & $1.18\pm0.15$ & $26.784\pm0.022$ & $3.1\pm0.5$ & $9.1\pm0.5$ & $79.2\pm1.5$ \\
            TS224959+105223 & $25.264\pm0.022$ & $-18.742\pm0.025$ & $-19.04\substack{+0.13\\-0.11}$ & $0.63\substack{+0.2\\-0.18}$ & $3.4\pm0.4$ & $25.324\pm0.022$ & $4.0\pm0.6$ & $11.7\pm0.6$ & $59.9\pm2.1$ \\
            TT231243+104321  & $25.22\pm0.04$ & $-15.36\substack{+0.06\\-0.05}$ & $-15.5\substack{+0.17\\-0.15}$ & $0.35\substack{+0.28\\-0.24}$ & $0.131\substack{+0.019\\-0.02}$ & $25.18\pm0.04$ & $0.78\pm0.19$ & $2.26\pm0.19$ & $12.4\pm0.7$ \\
            TT231241+104417 & $23.857\pm0.022$ & $-17.194\pm0.023$ & $-17.44\pm0.08$ & $0.54\substack{+0.13\\-0.12}$ & $0.79\pm0.06$ & $23.864\pm0.022$ & $0.98\pm0.19$ & $2.62\pm0.19$ & $15.8\pm0.7$ \\
            TT231329+102245 & $25.852\pm0.025$ & $-16.74\pm0.04$ & $-17.14\substack{+0.27\\-0.21}$ & $0.8\substack{+0.5\\-0.4}$ & $0.6\pm0.13$ & $26.056\pm0.026$ & $2.9\pm0.6$ & $8.3\pm0.6$ & $20.8\pm1.8$ \\
            TT231327+102227 & $25.362\pm0.022$ & $-17.683\pm0.024$ & $-18.02\substack{+0.16\\-0.14}$ & $0.7\substack{+0.27\\-0.24}$ & $1.35\pm0.19$ & $25.162\pm0.021$ & $2.4\pm0.6$ & $7.8\pm0.6$ & $35.9\pm1.8$ \\
            SA235044+271641 & $25.674\pm0.022$ & $-16.751\pm0.024$ & $-16.95\substack{+0.11\\-0.1}$ & $0.45\substack{+0.18\\-0.16}$ & $0.5\pm0.05$ & $25.553\pm0.022$ & $1.82\pm0.22$ & $5.57\pm0.22$ & $31.4\pm0.8$ \\
            TS025630+155522 & $25.636\pm0.02$ & $-17.166\pm0.017$ & - & - & - & - & $1.38\pm0.29$ & $3.81\pm0.29$ & $40.8\pm1.0$ \\

        \end{tabular}
\end{table*}

\subsection{Structural Parameters}
\begin{figure}
    \centering
    \includegraphics[width=\linewidth]{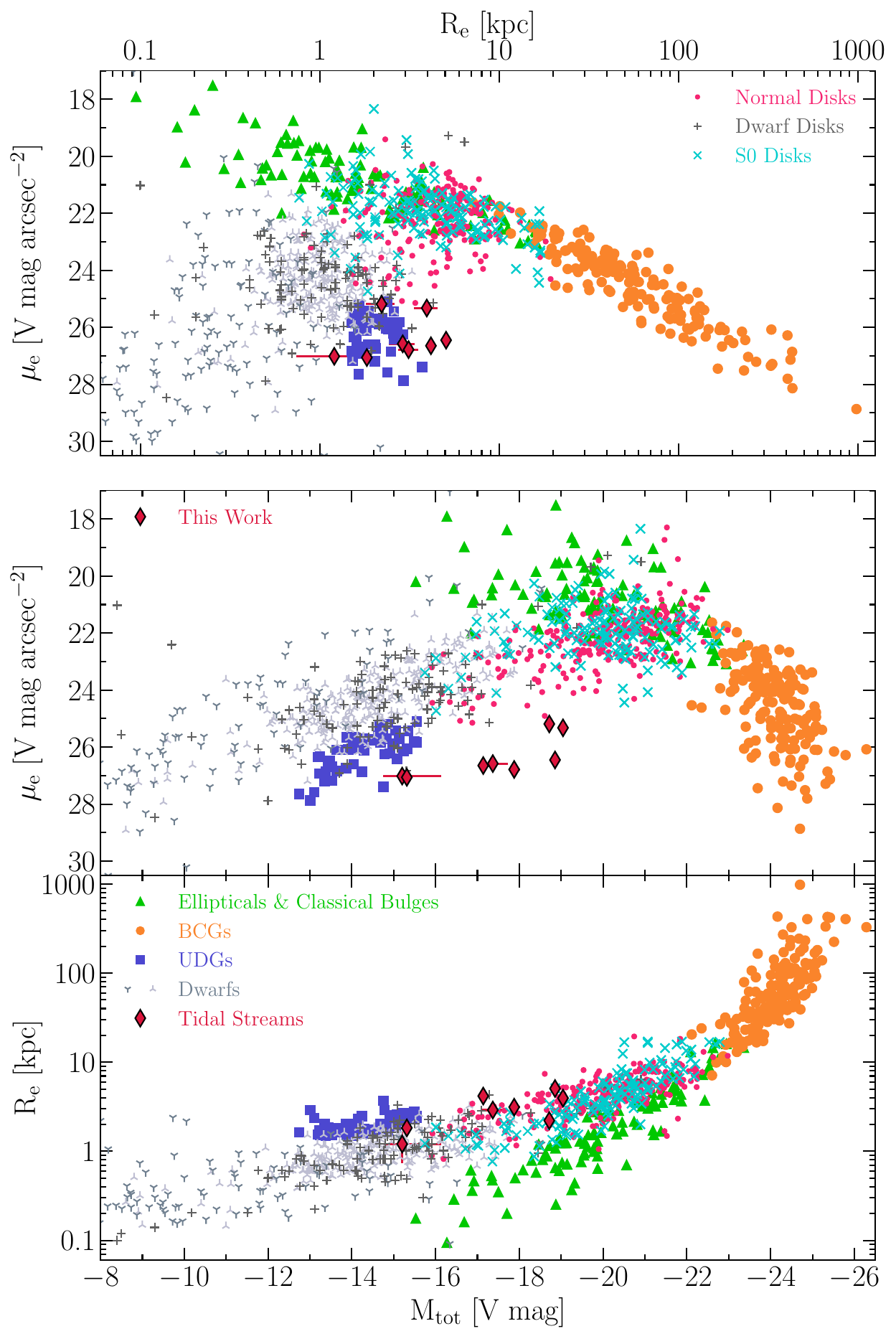}
    \caption{$R_\mathrm{e}-\mu_\mathrm{e}$, $M_\mathrm{tot}-\mu_\mathrm{e}$ and $M_\mathrm{tot}-R_\mathrm{e}$ parameter spaces of the tidal streams (red diamonds) in this work compared to dwarfs (light gray and slate gray)}, UDGs (blue), ellipticals (green), BCGs (orange), dwarf disks (dark gray), S0 disks (cyan), and normal disks (magenta). \protect\cite{kormendy+2009} delivered the basis for this figure with updates in \protect\cite{kormendy+2012}, \protect\cite{bender+2015}, and \protect\cite{kluge+2020}.
    \label{fig:structparams}
\end{figure}
In Figure \ref{fig:structparams}, the structural parameters, i.e., effective radius $R_\mathrm{e}$, effective surface brightness $\mu_\mathrm{e}$ and the total $V$-band brightness $M_\mathrm{tot}$, of eight out of the nine streams in our sample (TS025630+155522 had no complementing $r'$-band) are investigated. For reference, we show other various types of galaxies. The original figure is Figure 37 in \cite{kormendy+2009}. It was further updated by \cite{kormendy+2012}, \cite{bender+2015}, and \cite{kluge+2020}. Green points show the structural parameters for Ellipticals from \cite{bender+1992} and \cite{kormendy+2009} as well as classical bulges from \cite{fisher+2008}, \cite{kormendy+2009}, and \cite{kormendy+2012}. The orange data points are BCGs from \cite{kluge+2020} who used the same imaging dataset as in this work. Local group spheroidals are from \cite{mateo1998}, and \cite{mcconnachie+2006} and Virgo spheroidals from \cite{ferrarese+2006}, \cite{kormendy+2009}, and \cite{gavazzi+2015}, both shown in light gray. A catalog of newer local dwarfs, summarized in a database \citep{pace2024}, is shown in slate gray. They help to populate the faint regions. The A1656 UDG sample (blue squares) is from \cite{zoeller+2024}. Disks are grouped into dwarf disks \citep{pildis+1997,makarova1999,kirby+2008}, S0 disks \citep{baggett+1998,gavazzi+2000,laurikainen+2010,kormendy+2012}, and normal disks \citep{baggett+1998,gavazzi+2000,ferrarese+2006,laurikainen+2010,kormendy+2012}. A detailed figure of the disks can be found in \cite{kormendy+2012}. Red diamonds with a black outline display the stellar tidal streams studied in this work. While they are extended elongated objects, the classical definition of $R_\mathrm{e}$ does not hold. We measure the effective width along the minor axis, whereas the definition of $R_\mathrm{e}$ is along the major axis of an ellipse. However, we assume that when a satellite gets accreted, the effective radius does either not change or increase over time and is consistent with the effective width of the stream, i.e., the width of a Gaussian, within half of the total flux is captured. Furthermore, we must modify the effective width, since it is uncertain whether our measured effective width equates to the semi-major axis effective radius. A galaxy is also statistically unlikely to get stripped exactly parallel to its major or minor axes. Therefore, we assume that the measured effective width corresponds to the mean radius of an ellipse. Thus, we use $R_\mathrm{e}^\mathrm{major} = R_\mathrm{e}/\sqrt{b/a}$, where $R_\mathrm{e}$ is the effective radius and $R_\mathrm{e}^\mathrm{major}$ is the corrected effective width. We take the mean axis ratio of $b/a=0.76$ from the Coma cluster members in \cite{zoeller+2024}, which is consistent with the expected distribution of spheroid-shaped galaxies \citep{ryden1996,padilla+2008}.

The faint and elongated streams overlap with the UDGs in Figure \ref{fig:structparams} (top panel). However, the streams are brighter by a few magnitudes in total brightness $M_\mathrm{tot}$ (bottom panel). A similar trend is observed in the $M_\mathrm{tot}-R_\mathrm{tot}$ parameter space, where they shift toward the brighter end of the spheroidal population, predominantly aligning with S0 and normal disks.

\subsection{Coma Red Sequence}
\begin{figure}
    \centering
    \includegraphics[width=\linewidth]{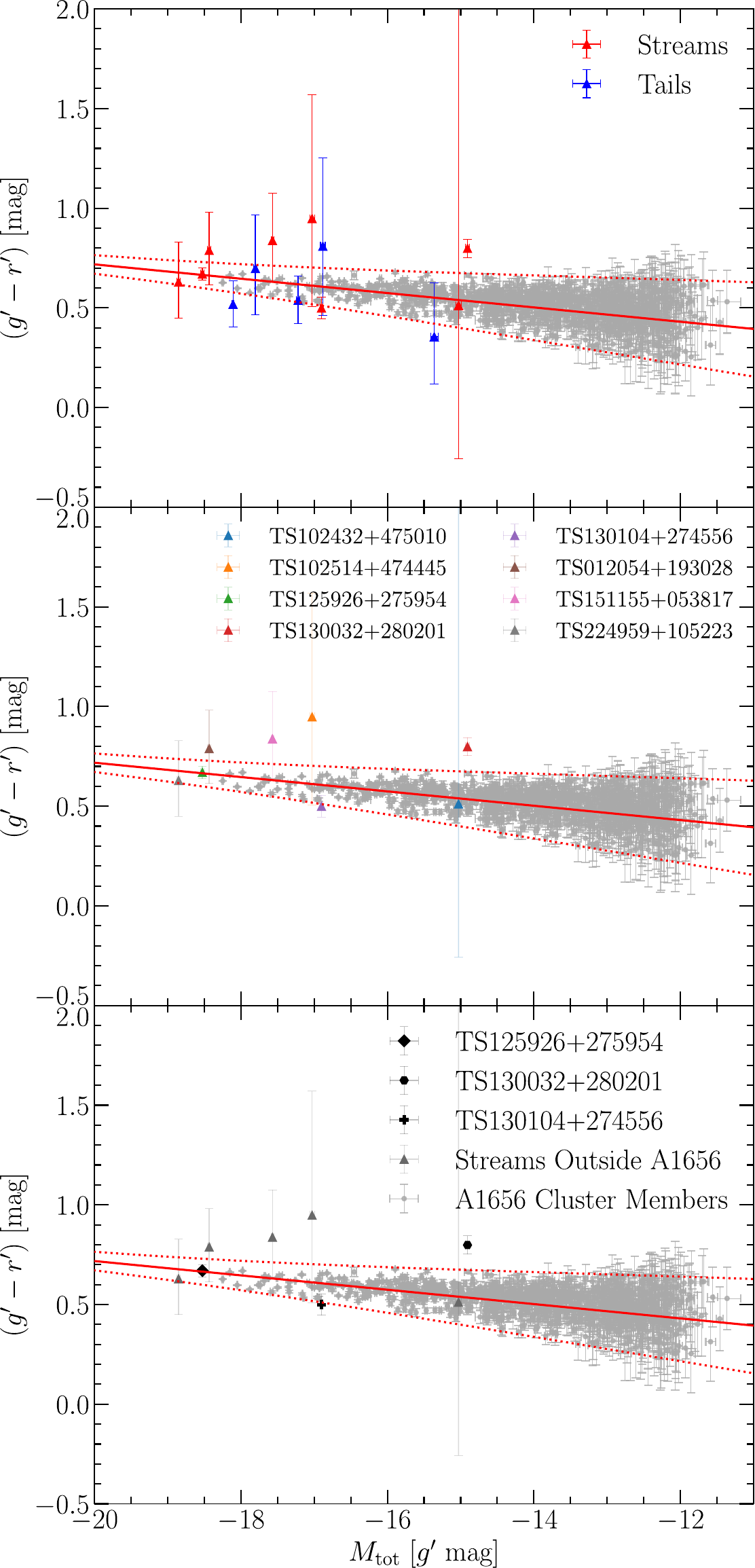}
    \caption{Tidal streams and tails compared to quiescent cluster member galaxies in the Coma cluster. Gray points with error bars are from the analysis in \cite{zoeller+2024}. Streams are shown in red, and tails are shown in blue with corresponding error bars. Only tidal streams in triangles with corresponding names are plotted in the middle panel.  Highlighted streams originally located in the Coma cluster with diamond, hexagonal, and cross-shaped markers are plotted in the lower panel. Triangles are the remaining streams.}
    \label{fig:redsequence}
\end{figure}

Another tool to study the nature of streams involves examining whether they align with the red sequence of galaxies (Figure \ref{fig:redsequence}). This might give an additional insight into what type of progenitor galaxy was stripped. Passive galaxies follow a red sequence because the optical colors barely evolve after an age of 9 Gyr \citep[e.g.,][]{bruzal+2003}. The tilt can be explained by the mass-metallicity relation \citep[e.g.,][]{gu+2020}. In scenarios where star formation is absent, often in dry minor mergers, the colors and total luminosity remain unchanged. Therefore, the properties of streams, which may originate from early-type galaxies (ETGs), are expected to align with the red sequence. To achieve this, we utilize the red sequence data of the Coma cluster \citep{zoeller+2024} and compute rest-frame colors by applying $K$ corrections and correcting the total brightnesses with the distance modulus.

The result is shown in Figure \ref{fig:redsequence}. In the upper panel, both streams (in red) and tails (in blue) are plotted in the color-magnitude space. Both distributions occupy the brighter end, albeit exhibiting considerable dispersion around the narrow limits of the red sequence, due to the high measurement uncertainties. The middle panel focuses solely on the streams within our sample. The error bars are the same as in the top panel but not shown, as only the positions of the streams on the red sequence matter here. On the whole, streams align with the red sequence within $2\sigma$.

The bottom panel highlights the three streams within the Coma cluster. While those located outside the cluster are calibrated solely to the Coma distance, we scrutinize their positioning on the red sequence. The largest stream TS125926+275954 demonstrates good agreement with the red sequence, which is unsurprising given its relatively large apparent size, and luminosity, and visual color. Similarly, tails are also located at the bright end of the red sequence within a $1\sigma$ confidence. They, however, show a slight trend toward bluer colors than the streams do.
\section{Discussion}\label{sec:discussion}
\subsection{Progenitors of Streams}
We undertake a comparative analysis of the structural parameters of tidal streams against those of ellipticals, BCGs, spheroidals, disks, and UDGs to elucidate their positioning within the depicted parameter spaces (Figure \ref{fig:structparams}). This comparison is needed to infer possible progenitor candidates for the streams in our sample.

There are three main properties we want to consider in this discussion: (1) The total brightness, (2) the color (also see Section \ref{subsec:redseq}), and (3) the size, i.e., effective radius (also see Section \ref{subsec:reff}). Those strongly depend on the age of the stream, i.e., the time passed since the first stripping of stars. The total brightness further depends on star formation in the stream. If star formation is triggered by the tidal forces, bright massive stars form, and this would increase the total brightness of the stream, even long after their formation. \textbf{The color would not be strongly affected if the stream is older than 1 Gyr \citep{bruzal+2003}, as those stars quickly move to the red giant or supergiant branch}. In addition, the stream gets redder due to the aging of its stellar population, i.e., the population of its progenitor.

Beforehand, we split our stream sample into three categories based on the definition of dwarf galaxies as they are most commonly associated as tidal stream progenitors. The dwarf regime is generally defined as fainter than $M_V=-17$ mag \citep[e.g.,][]{tammann1994} or sometimes with a more conservative approach of $M_V\gtrsim-18$ \citep[e.g.][]{grebel+2003}. It follows that we have two faint streams (TS102432+475010, TS130032+280201), with $M_V$ fainter than $-17$ mag, three "standard" streams (TS102514+474445, TS130104+274556, TS151155+053817), with $-18\leq M_V\leq-17$ and three bright streams (TS012054+193028, TS224959+105223, TS125926+275954) with $M_V\leq-18$.
\begin{center}
   \textit{BCGs}
\end{center}
Generally, BCGs are ruled out as progenitors. Given their status as the brightest and most-massive galaxies in the Universe, any interaction involving a BCG as the infalling less-massive merger member would still result in a major merger rather than a tidal stripping event. Therefore, BCGs are generally not considered as plausible progenitors for tidal streams.
\begin{center}
    \textit{Dwarfs}
\end{center}
To reach the luminosity of our bright streams, progenitors from the dwarf population need to increase their flux by up to a factor of $2.5$, even when using the conservative definition of $M_{V,\mathrm{Dwarf}} \geq -18$. For those, a dwarf progenitor is unlikely, whereas for the faint and standard streams, dwarfs are a valid progenitor candidate. We justify this claim, as the bright streams are still connected to their progenitor. As a lot of flux is still present in the progenitor nucleus, the true total brightness would increase significantly. Additionally, we observe them early in their formation, which would mean that it is likely that newborn stars make them bluer. However, we observe red colors, namely $0.79$, $0.63$, and $0.67$, respectively. We conclude both arguments make a (conservative) dwarf even more unlikely. For the large stream in the Coma cluster, i.e., TS125926+275954, it also possible that the stream has two progenitors, due to its strong asymmetric shape and two overlapping galaxies with similar width.

For the faint streams, dwarfs are the only possible explanation, because they are fainter than the faint end of the disk, S0 and E population. Also, the size arguments holds, as  the faint streams populate the bright-large end of the dwarf population.

A dwarf progenitor example from the literature comes from \cite{foster+2014}, who studied the stream around the Umbrella galaxy, NGC 4651, and measured a total brightness of $M_V=-17.0$, which is consistent with that of dwarf galaxies. They further assume an effective radius of $R_\mathrm{e}\sim1$kpc and label the progenitor as a dwarf elliptical. This measurement agrees well with our stream sample in the bottom panel of Figure \ref{fig:structparams}.

\begin{center}
    \textit{Disks, Ellipticals, or S0s}
\end{center}
We further investigate S0 and normal disks, as well as ellipticals as other potential progenitors. Among the bright streams, S0 or ellipticals are favored due to their gas loss and low to negligible star formation rates, consistent with the observed red colors of the streams (see \ref{fig:redsequence}). Distinguishing from S0s or Es as progenitors is difficult due to the behavior of the effective width of the stream. If $R_\mathrm{e}$ would increase, the stream moves upward in the lower panel of Figure \ref{fig:structparams}, which would explain their origin from ETGs. The other scenario involves a nonchanging $R_\mathrm{e}$ and hence would suggest S0 progenitors. This leaves both as plausible candidates for the origins of these three streams. A study by \cite{errani+2015} shows increasing width, but this depends on the time at which we observe the stream. Therefore, we cannot claim either S0 or ETGs progenitors.

We apply the same arguments for the standard streams, as they show similar brightnesses as the faint end of the disk, S0 and E population. The difference is that for those, dwarfs can also be considered.

Both the $M_\mathrm{tot}-\mu_\mathrm{e}$ and $R_\mathrm{e}-\mu_\mathrm{e}$ parameter spaces have the same implication. Streams move downward in the middle panel of Figure \ref{fig:structparams} because their area increases simply from tearing apart the progenitor galaxy while the flux is conserved. Lastly, the upper panel shows both arguments for $\mu_\mathrm{e}$ and $R_\mathrm{e}$. Namely, with an increased area and constant brightness, the surface brightness decreases. Again, we cannot imply any evolution of $R_\mathrm{e}$ in this panel. Both arguments are also valid if the brightness would increase due to star formation. The descent is then diagonal, but still does not exclude any of the progenitor types discussed before.

Additionally, we estimate the age of the streams from typical velocities of the infalling progenitor, i.e., the velocity dispersion of the cluster ($\sim1000$km s$^{-1}$), and their projected length. Our streams are between $10^8$ and $5\times10^8$ yr old, which is only a rough estimate! However, as the streams are also red (Figure \ref{fig:redsequence}), triggered star formation by the merger is unlikely, as it would make them bluer. This circumstance opens the possibility for dwarf ellipticals (dEs) as progenitors for the "standard" and bright streams in our sample, as their total brightness would remain constant while the size increases.

\begin{figure}
    \centering
    \includegraphics[width=\linewidth]{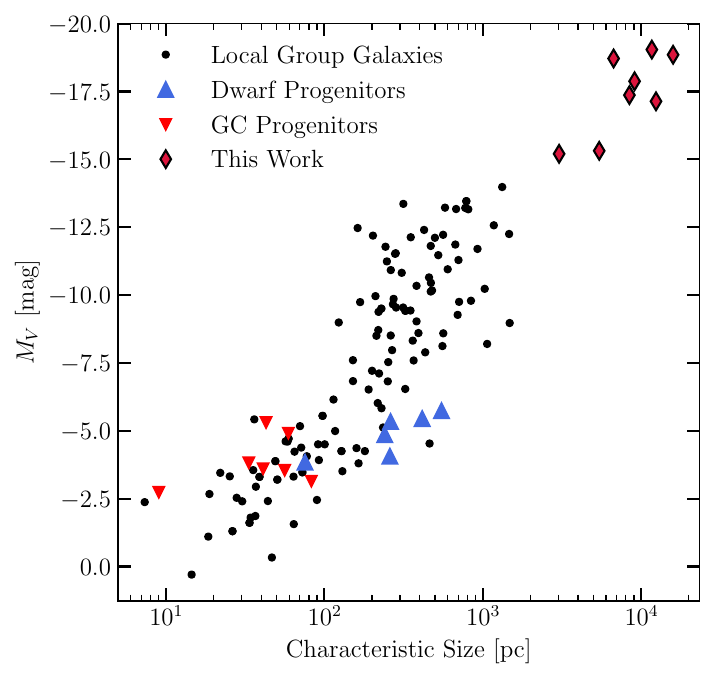}
    \caption{Characteristic size of tidal streams (triangles and diamonds) and known Local Group galaxies (black circles) against their total absolute $V$-band magnitude. The literature data (Local Group galaxies, GC progenitors, and dwarf progenitors) are taken from \cite{patrick+2022}.}
    \label{fig:patrickstreams}
\end{figure}

Finally, we compare our streams to stream progenitors and Local Group dwarfs \cite[see also Figure 40 in][]{patrick+2022} in Figure \ref{fig:patrickstreams}. They analyzed streams in the galactic halo and measured their characteristic size, i.e., average width along the stream, and their absolute magnitude and linked them dwarf and GC progenitors. Our streams extend the distribution to the bright and wider end. We apply the same brightness argument, where in this Figure, for $75\%$ of the streams, a local dwarf progenitor seems very unlikely, even if star formation would be induced during the stripping.

\subsection{Connecting effective radius and width}\label{subsec:reff}
The effective radius provides the remaining distinguishing factor between the (dwarf) Es and S0s. It is worth noting that the measurement of the effective width might be subject to temporal variability. That means the effective width of the stream depends on the current stage of the merger and the satellite stripping.

We do not know how exactly $R_\mathrm{e}$ behaves during and after the minor merger. There is an analysis by \cite{errani+2015}, which uses high-resolution $N$-body simulations to trace the formation and evolution of tidal streams across multiple pericentric passages coming from dwarf spheroidal galaxies. They found that the half-light radius of a stream increases with distance from the host galaxy based on the stripped particles at the pericenter. Generally this result supports our assumption of a constant or increasing $R_\mathrm{e}$ for this discussion. However, for our discussion, this also depends on the stripping stage of the stream, which we do not know. Hence, we do not discard the assumption of equal width for the discussion on possible progenitors.

In addition, simulations like the Illustris TNG50 \citep{nelson+2019} predict the merging of lower-mass satellites with a larger host, while their stellar content gets stripped in the process. The emergence of multiple stellar streams is visible throughout the simulation. However, how the effective radius of the progenitor evolves during and after the stripping has not been studied to the extent that it is needed. We suspect that it depends on multiple variables, such as impact parameter, relative velocities, size, mass, position angle with respect to the major and minor axis radii, and the impact parameter.

\subsection{Streams and Tails on the Red Sequence}\label{subsec:redseq}

For Tails, we neglect star-forming regions in our measurements. Prior to modeling, these regions are either masked or interpolated. As we measure fluxes, those missing pixels are replaced by the model. Given that newly formed stars tend to be bright and blue, the true nature of the tails in Figure \ref{fig:redsequence} is expected to be brighter and bluer. The magnitude of this effect depends on factors such as the number and size of star-forming regions. We do not quantify this effect. However, it was also shown in \cite{mulia+2015} that tails have diverse color ranges from red to blue colors depending on their host and the type of major merger.

All streams align with the red sequence of quiescent cluster members within a 2$\sigma$ uncertainty. Those galaxies don't form stars by definition. While this would suggest that the streams in our sample originate from non-star-forming cluster members, the large error bars, especially in the color measurements from the LS, are also consistent with star-forming progenitors, or, perhaps, streams with triggered star formation caused by tidal forces. An improvement to distinguish between a passive and an active stream would be an analysis of the $u'-g'$ color. Unfortunately, the LS did not provide $u$-band observations. The WWFI Coma streams are covered by the $u'$-band but, there, the low S/N becomes increasingly more critical and makes the final $u'-g'$ color untrustworthy. An in-depth analysis is beyond the scope of this work.

In addition, we investigated the influence of the color in different aperture segments. We found that the scatter is consistent with the Poisson noise and background variations. The uncertainty of the local background has the strongest impact on the brightness measurements. We calculate the local background from a weighted average of the fitted Gaussian offsets. This average, however, is not able to capture variations in the background, e.g., in dense regions as for TS125926+275954 in the center of the coma cluster. For large streams with a lot of pixels, a small change in the choice of the local background can shift the colors of the feature by $0.1$ mag. Other methods, such as background modeling and removal in inhomogeneous regions were not applied, and hence might improve those measurements, especially for those of the three streams in the Coma cluster.

\section{Summary}\label{sec:summary}
We have measured the properties of tidal streams and tails detected and classified in the Abell cluster sample in \citep{kluge+2020,kluge+2021}. We utilize the DR10 from the Legacy Survey to complement our $g'$-band with $g$ and $r$ bands to obtain color information for each feature. A novel algorithm was developed to model these features and to create robust apertures, which were used for photometric analysis. The outputs of the stream modeling can be found in Appendix \ref{app:outputs}, Figure \ref{fig:chap-results:ts0120+1930} to \ref{fig:app:streams:end}. Our main findings are as follows:

\begin{enumerate}
\item Dwarf, lenticular, and \textbf{(dwarf)} elliptical galaxies are the proposed \textbf{progenitor} candidates for the streams in this work. For the three brightest streams, a progenitor of the latter two is more likely due to their brightness, which is a factor of 2 larger than the conservative definition of dwarf galaxies.
\item Streams are on average dimmer ($\overline{SB}\approx26.25$ $g'$ mag arcsec$^{-2}$) and redder ($\overline{(g'-r')}\approx0.70$ mag) than tails ($\overline{SB}\approx25.14$ $g'$ mag arcsec$^{-2}$ and $\overline{(g'-r')}\approx0.56$). This is in good agreement with the general classification and properties of major (tails) and minor mergers (streams).
\item We find that streams show sub-Milky Way luminosities with an average of $1.5\times10^{9}\,L_\odot$.
\item The analyzed streams follow the red sequence of the Coma cluster within $2\sigma$ while located on the bright end relative to Coma member galaxies.
\item Tidal tails follow the same structural description as streams; namely, both can be described by 1D Gaussians with higher-order moments, perpendicular to their direction. At least, this is true for their faint underlying component.
\item Both streams and tails are efficiently modeled with our algorithm and with high accuracy of $7\%$ for brightnesses or lower and $3\%$ for the shape, regardless of the complexity of the stream.
\end{enumerate}

\section{Acknowledgments}
We thank the anonymous referee for providing comments and suggestions that allowed us to significantly improve the paper.
 
The Wendelstein 2.1\,m telescope project was funded by the Bavarian government and by the German Federal government through a common funding process. Part of the 2.1\,m instrumentation including some of the upgrades for the infrastructure were funded by the Cluster of Excellence “Origin of the Universe” of the German Science foundation DFG.

This paper uses data that were obtained by The Legacy Surveys: the Dark Energy Camera Legacy Survey (DECaLS; NOAO Proposal ID \# 2014B-0404; PIs: David Schlegel and Arjun Dey), the Beijing-Arizona Sky Survey (BASS; NOAO Proposal ID \# 2015A-0801; PIs: Zhou Xu and Xiaohui Fan), and the Mayall z-band Legacy Survey (MzLS; NOAO Proposal ID  \# 2016A-0453; PI: Arjun Dey). DECaLS, BASS, and MzLS together include data obtained, respectively, at the Blanco telescope, Cerro Tololo Inter-American Observatory, National Optical Astronomy Observatory (NOAO); the Bok telescope, Steward Observatory, University of Arizona; and the Mayall telescope, Kitt Peak National Observatory, NOAO. NOAO is operated by the Association of Universities for Research in Astronomy (AURA) under a cooperative agreement with the National Science Foundation. Please see \url{http://legacysurvey.org} for details regarding the Legacy Surveys. BASS is a key project of the Telescope Access Program (TAP), which has been funded by the National Astronomical Observatories of China, the Chinese Academy of Sciences (the Strategic Priority Research Program "The Emergence of Cosmological Structures" Grant No. XDB09000000), and the Special Fund for Astronomy from the Ministry of Finance. The BASS is also supported by the External Cooperation Program of Chinese Academy of Sciences (Grant No. 114A11KYSB20160057) and Chinese National Natural Science Foundation (Grant No. 11433005). The Legacy Surveys imaging of the DESI footprint is supported by the Director, Office of Science, Office of High Energy Physics of the U.S. Department of Energy under Contract No. DE-AC02-05CH1123, and by the National Energy Research Scientific Computing Center, a DOE Office of Science User Facility under the same contract; and by the U.S. National Science Foundation, Division of Astronomical Sciences under Contract No.AST-0950945 to the National Optical Astronomy Observatory.

\textit{Software:} Astropy \citep{astropy2022}, Photutils \citep{bradley+2023}, numpy \citep{harris+2020}, SciPy \citep{scipy}

\bibliography{paper}{}
\bibliographystyle{aasjournal}

\appendix
\section{Modeling Algorithm}

We conducted extensive testing to evaluate the performance and quality of our modeling algorithm. Owing to its low computational cost, our algorithm is well suited for automated pipelines handling large datasets. Although some manual intervention is required, these tasks can also be automated. For instance, the selection of the position and size of the initial box could be trained using a neural network, which may already be employed to identify such features in vast fields. Alternatively, it could be used to determine the centroid of a polygon-like segmentation mask, also provided by a neural network.

Our algorithm does have its limitations. Features with pronounced curvature in a projected view, particularly those observed at low inclination angles, cannot be effectively modeled. This is because we treat the angle fit parameter as to be constant in the cutout feature segment. For parts with strong curvature, this assumption does not hold. Room for improvement lies in the behavior of the modeling box. Currently, it remains either vertical or horizontal, whereas employing a rotating box could enhance the modeling process by ensuring that the box remains perpendicular to the orientation of the feature.

We also emphasize that our intrinsic best-fit parameters are those for an assumed circular Gaussian PSF. However, as asymmetric PSFs exist, and PSFs are often described by other forms, e.g., Moffat profile, using the true PSF of the image would result in the true intrinsic shape of the feature. Although, we suspect that this change will only slightly change the fit parameters.

\section{The Gaussian Box Framework}\label{app:gaussbox}
In the following, we describe how a Gaussian is fitted to a 2D finite box with a width $W\equiv2w+1$ and a height $H\equiv2h+1$. Both can be expressed in terms of ranges, for example, the width ranges from $-h$ to $h$. With that, we construct two vectors $\Vec{x}$ and $\Vec{y}$ representing the horizontal and vertical dimensions of the box, respectively.
\begin{equation}
    \Vec{x}(w)=
        \begin{bmatrix}
            -w & \dots & w
        \end{bmatrix}
    ,\,
    \Vec{y}(h)=
        \begin{bmatrix}
            h\\
            \vdots\\ 
            -h
        \end{bmatrix}
    \,
    w,h\in\mathbb{Z}
\end{equation}
Not only do they have the same corresponding dimensions for the sides of the box, but encapsulate distance information, i.e., each value in these vectors directly represents the pixel distance from the center. When stacking $\Vec{x}$ vertically and $\Vec{y}$ horizontally, we end up with two matrices with the same dimensions as the box ($\dim(\mathbf{x})=\dim(\mathbf{y})=2h+1\times2w+1$).
\begin{align}
    \mathbf{x}=\text{vstack}\{\Vec{x}\}=
    \begin{bmatrix}
        -w&\dots&w\\
        \vdots&\ddots&\vdots\\
        -w&\dots &w
    \end{bmatrix}\\
    \mathbf{y}=\text{hstack}\{\Vec{y}\}=
    \begin{bmatrix}
        h&\dots&h\\
        \vdots&\ddots&\vdots\\
        -h&\dots&-h
    \end{bmatrix}
\end{align}
By adding both matrices together, we get a distance grid, where each pixel stores the information of its distance to a certain "zeroth" line. The idea behind it is that when feeding the grid into a Gaussian function, the "zeroth" line transforms to values of one, while the rest distributes themselves symmetrically around the mean (when taking a 1D slice).
\begin{equation}
    \mathbf{\Delta} = \mathbf{x} + \mathbf{y} = \begin{bmatrix}
        -w + h & \dots & w + h\\
        \vdots & \ddots & \vdots\\
        -w - h & \dots & w - h
    \end{bmatrix}
\end{equation}
To allow for a grid rotation and translation, the simple addition is modified by angle and offset parameters. Our grid is now constructed so that an angle of zero translates into a fully vertical stream segment with its orientation angle increasing counterclockwise. 
\begin{equation}
    \mathbf{\Delta}(\mathbf{x},\mathbf{y},x_0,y_0,\alpha)=\cos(\alpha)(x-x_0)+\sin(\alpha)(y-y_0)
\end{equation}
To create a stream segment, the grid is inserted into a Gaussian function. As \textbf{described in Section \ref{subsubsec:gaussianfit}} the Gaussian is extended to higher-order moments utilizing the Hermite basis functions and a cumulative distribution function. With this setup, every imaginable stream segment can be constructed with a total of nine fit parameters. They vary along the stream track and build up the entire structural description of the stream (or tail).

\section{76-98-99.9 Rule}\label{app:rule}
Here we derive the analog to the 68–95–99.7 rule for the FWHM criterion. The standard deviation and the FWHM are connected via
\begin{equation}
    \text{FWHM} = 2 \sqrt{2\ln2}\sigma \approx 2.3548\sigma.
\end{equation}
But the FWHM describes, as the name suggests, the full width; hence, the direct transformation from $\sigma$ only needs the half-width at half-maximum (HWHM).
\begin{equation}
    \text{HWHM} \equiv \lambda = \sqrt{2\ln2}\sigma \approx 1.1774\sigma.
\end{equation}
To find the percentage of the full area under a Gaussian distribution, we first assume a normalized Gaussian.
\begin{equation}
    P(x,\mu,\sigma) = \frac{1}{\sqrt{2\pi}\sigma} \exp\Biggl({-\frac{1}{2}\Bigl(\frac{x-\mu}{\sigma}\Bigr)^2}\Biggr)
\end{equation}
Now we integrate the function inside the borders.
\begin{equation}
    \frac{1}{\sqrt{2\pi}\sigma} \int_{\mu-n\sigma}^{\mu+n\sigma} \exp\Biggl({-\frac{1}{2}\Bigl(\frac{x-\mu}{\sigma}\Bigr)^2}\Biggr) dx
\end{equation}
By substituting $u = \frac{x-\mu}{\sigma}$, along $dx = \frac{1}{\sigma} du$ and $\mu\pm n\sigma \Rightarrow \pm n$  the integral simplifies to
\begin{equation}
    \frac{1}{\sqrt{2\pi}} \int_{-n}^{n} \exp\Biggl({-\frac{1}{2}u^2}\Biggr) du .
\end{equation}
Since we do not want to integrate from $-n$ to $n$, which would give us the area in steps of $\sigma$, we have to further substitute our definition for it, i.e., multiplying $\pm n$ by $\sqrt{2\ln{2}}$.
\begin{equation}
    \frac{1}{\sqrt{2\pi}} \int_{-\sqrt{2\ln{2}}n}^{\sqrt{2\ln{2}}n} \exp\Biggl({-\frac{1}{2}u^2}\Biggr) du .
\end{equation}
At the end, choosing $n\in\{1,2,3\}$ we arrive at
\begin{align}
    P(\mu - 1 \lambda \leq X \leq \mu + 1 \lambda) &\approx 0.7610\\
    P(\mu - 2 \lambda \leq X \leq \mu + 2 \lambda) &\approx 0.9815\\
    P(\mu - 3 \lambda \leq X \leq \mu + 3 \lambda) &\approx 0.9996
\end{align}
forming the 76–98–99.9 rule, where $\lambda$ is the HWHM.

\section{Stream and Tail Models}\label{app:outputs}
\begin{figure}
    \centering
    \includegraphics[width=\linewidth]{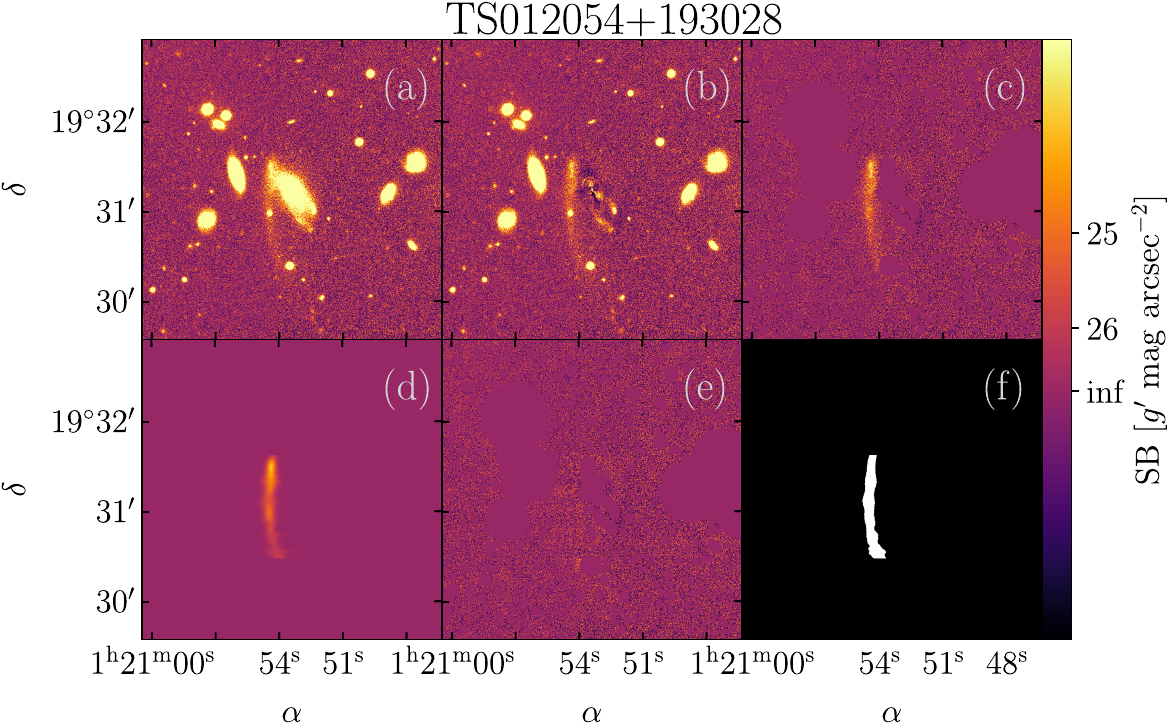}
    \caption{Visualization of the TS012054+193028 stream. (a) The original image cutout. (b) Image after subtracting an isophotal galaxy model. (c) \textbf{Source mask} multiplied onto panel (b). (d) The model created by \textsc{astrostreampy}. (e) The residual image. (f) The $\mu_1$ aperture.}
    \label{fig:chap-results:ts0120+1930}
\end{figure}

We provide the streams, their models, the masks, the residuals, and the $\mu_1$ apertures in Figure \ref{fig:chap-results:ts0120+1930} to \ref{fig:app:streams:end}.
\begin{figure}
    \centering
    \includegraphics[width=\linewidth]{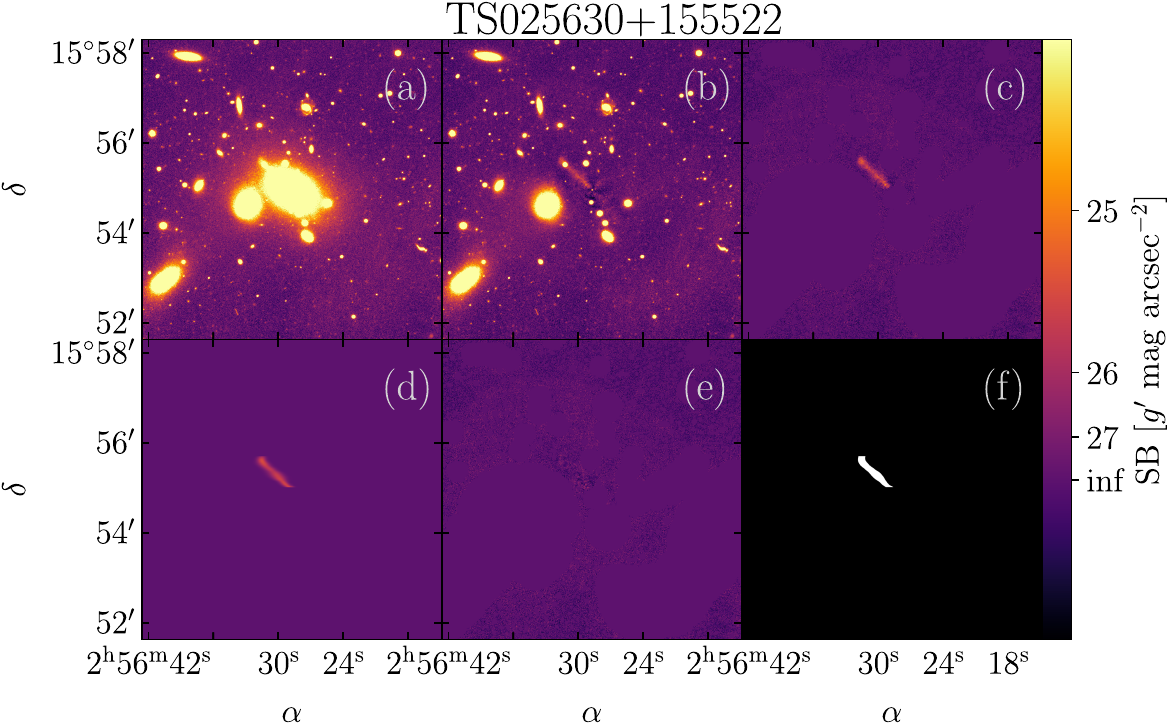}
    \caption{Same as in \hyperref[fig:chap-results:ts0120+1930]{Figure \ref{fig:chap-results:ts0120+1930}} but for the TS025630+155522 stream.}
\end{figure}
\begin{figure}
    \centering
    \includegraphics[width=\linewidth]{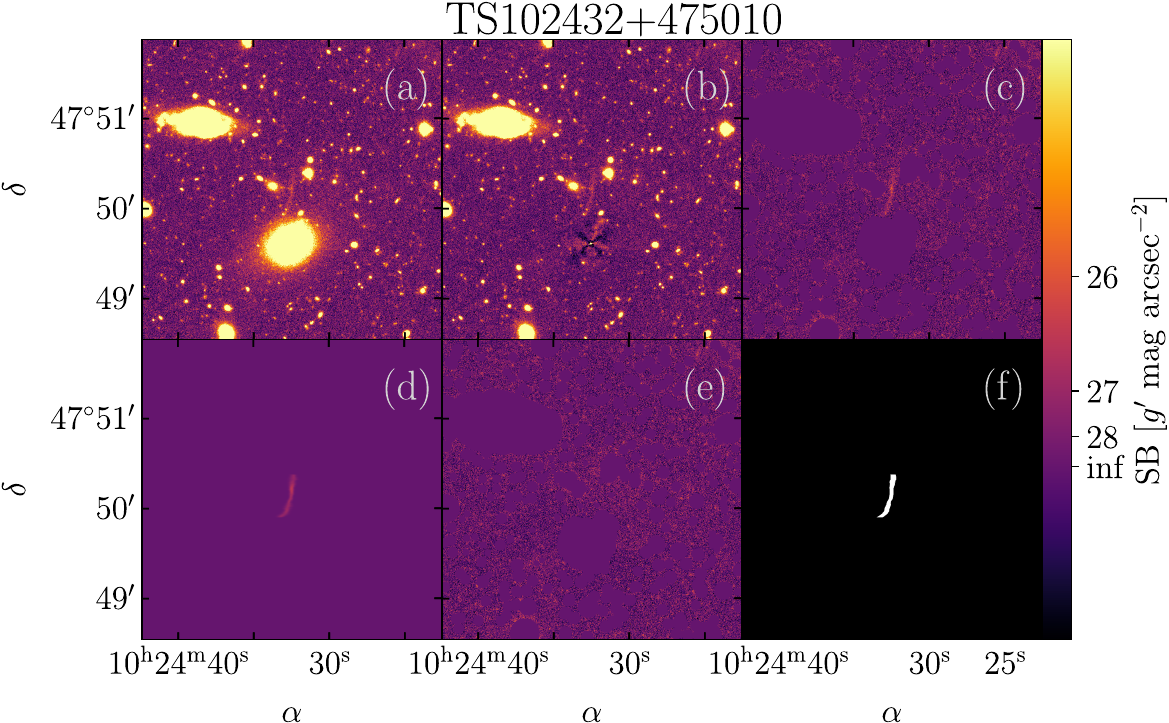}
    \caption{Same as in \hyperref[fig:chap-results:ts0120+1930]{Figure \ref{fig:chap-results:ts0120+1930}} but for the TS102432+475010 stream.}
\end{figure}
\begin{figure}
    \centering
    \includegraphics[width=\linewidth]{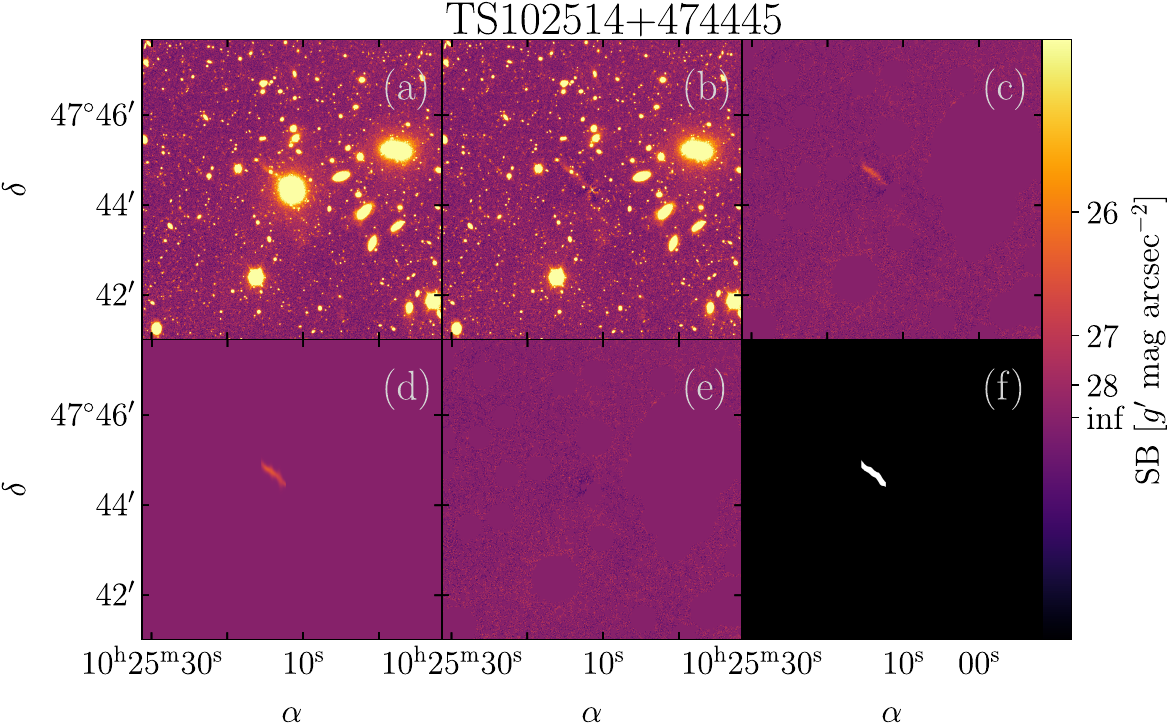}
    \caption{Same as in \hyperref[fig:chap-results:ts0120+1930]{Figure \ref{fig:chap-results:ts0120+1930}} but for the TS102514+474445 stream.}
\end{figure}
\begin{figure}
    \centering
    \includegraphics[width=\linewidth]{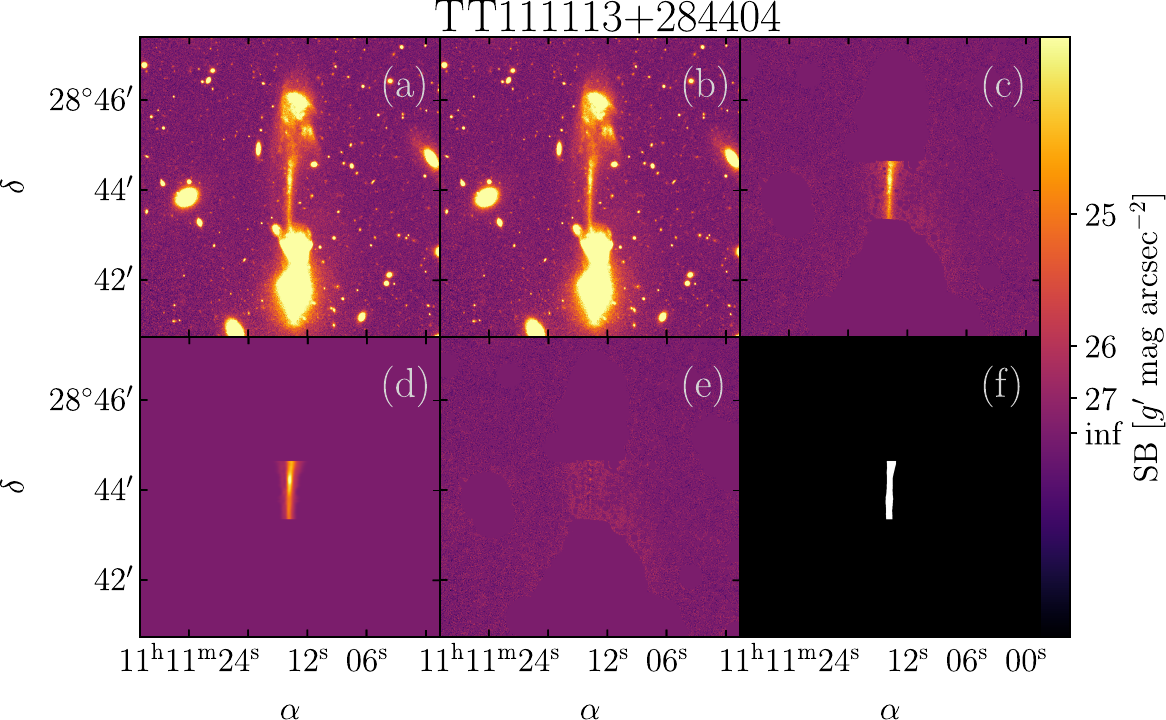}
    \caption{Same as in \hyperref[fig:chap-results:ts0120+1930]{Figure \ref{fig:chap-results:ts0120+1930}} but for the TT111113+284404 tail and without subtracting an isophotal model.}
\end{figure}
\begin{figure}
    \centering
    \includegraphics[width=\linewidth]{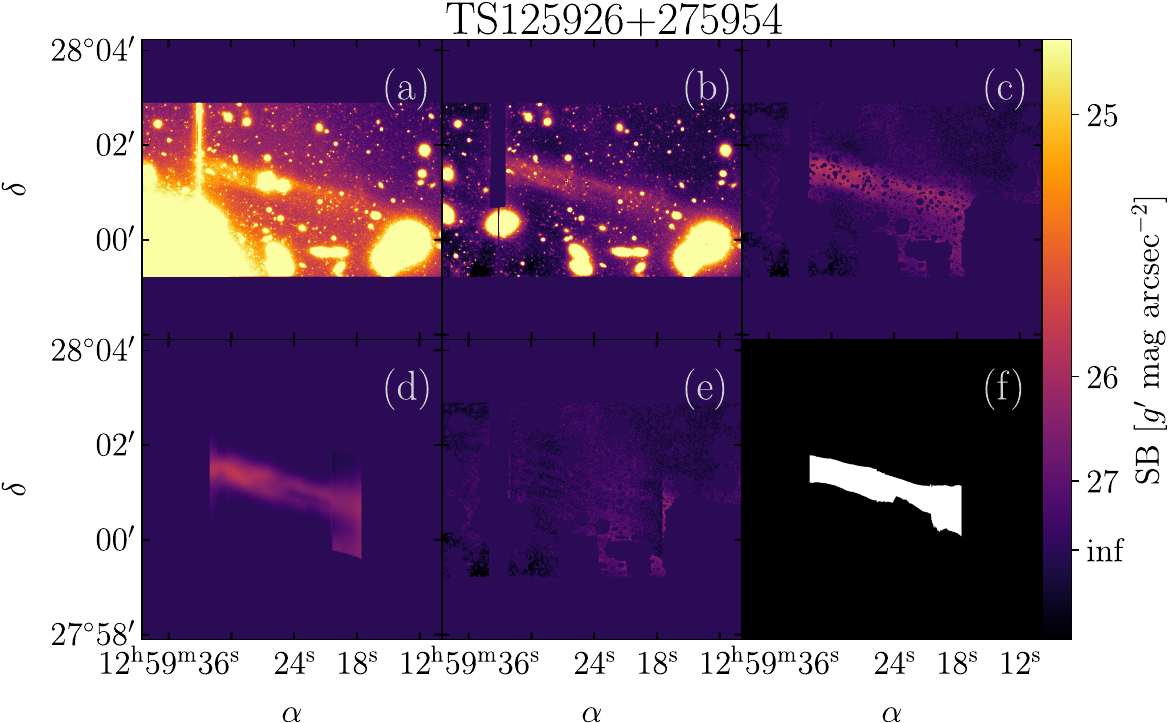}
    \caption{Same as in \hyperref[fig:chap-results:ts0120+1930]{Figure \ref{fig:chap-results:ts0120+1930}} but for the TS125926+275954 stream and with subtracting four galaxy models in total. Two of those are the BCGs not visible anymore in the final cutout.}
\end{figure}
\begin{figure}
    \centering
    \includegraphics[width=\linewidth]{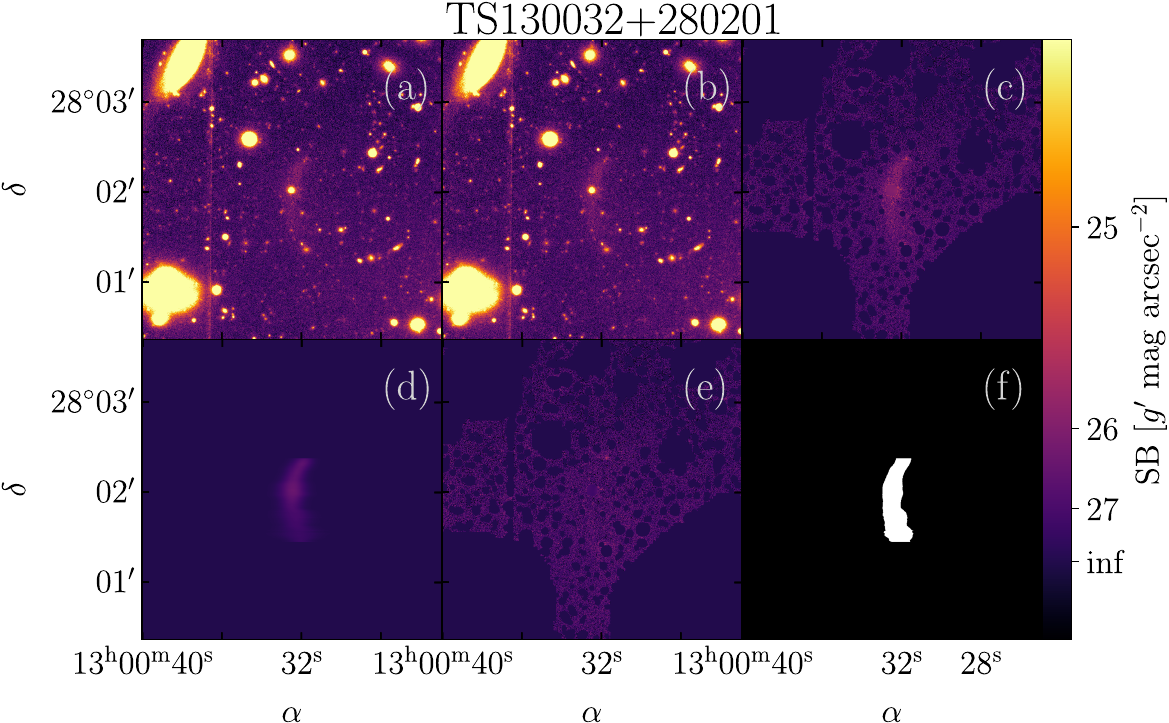}
    \caption{Same as in \hyperref[fig:chap-results:ts0120+1930]{Figure \ref{fig:chap-results:ts0120+1930}} but for the TS130032+280201 stream and without subtracting an isophotal model.}
\end{figure}
\begin{figure}
    \centering
    \includegraphics[width=\linewidth]{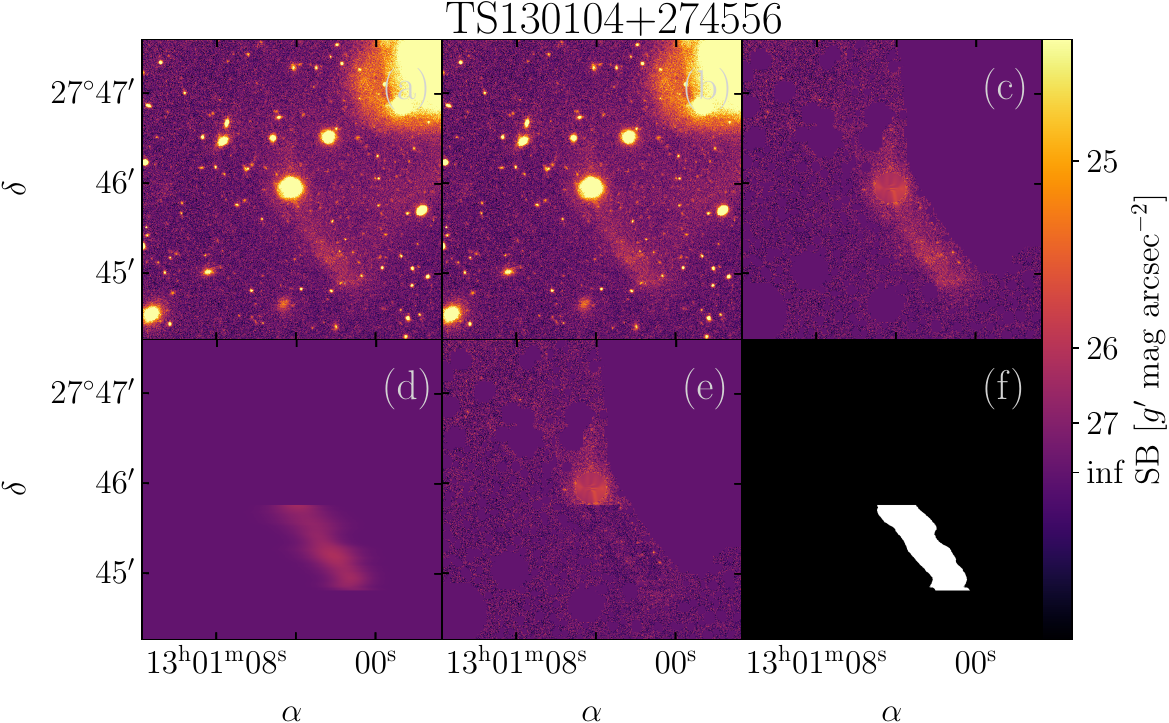}
    \caption{Same as in \hyperref[fig:chap-results:ts0120+1930]{Figure \ref{fig:chap-results:ts0120+1930}} but for the TS130104+274556 stream and without subtracting an isophotal model.}
\end{figure}
\begin{figure}
    \centering
    \includegraphics[width=\linewidth]{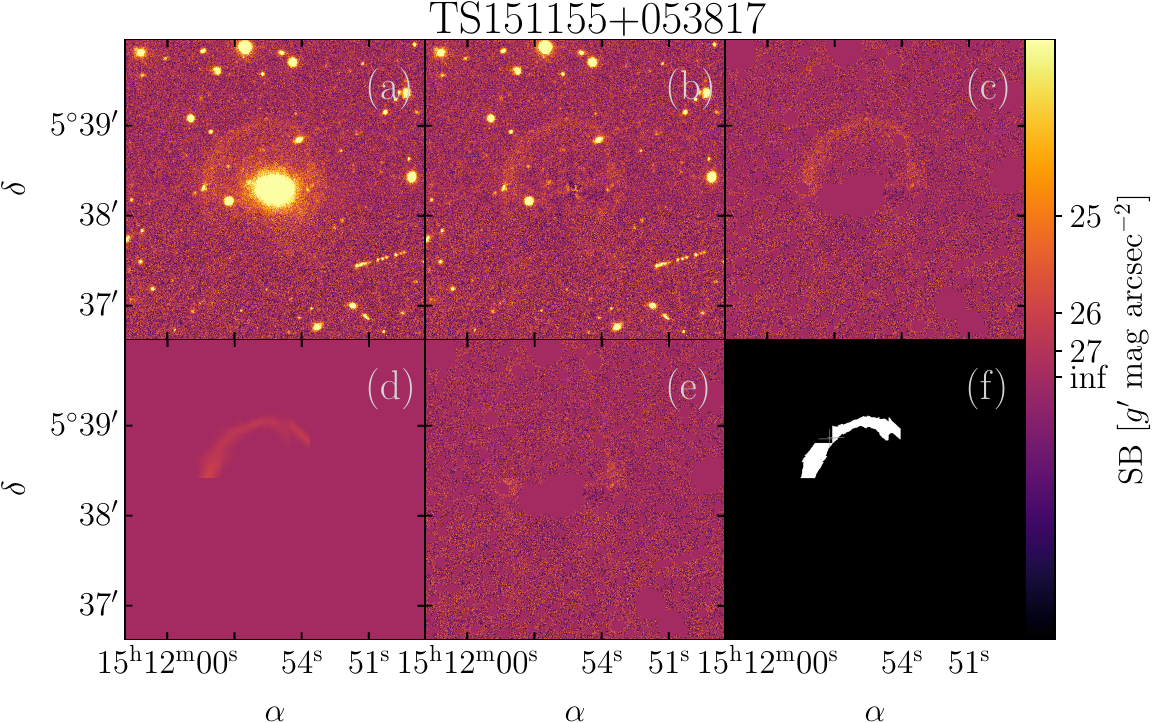}
    \caption{Same as in \hyperref[fig:chap-results:ts0120+1930]{Figure \ref{fig:chap-results:ts0120+1930}} but for the TS151155+053817 stream.}
\end{figure}
\begin{figure}
    \centering
    \includegraphics[width=\linewidth]{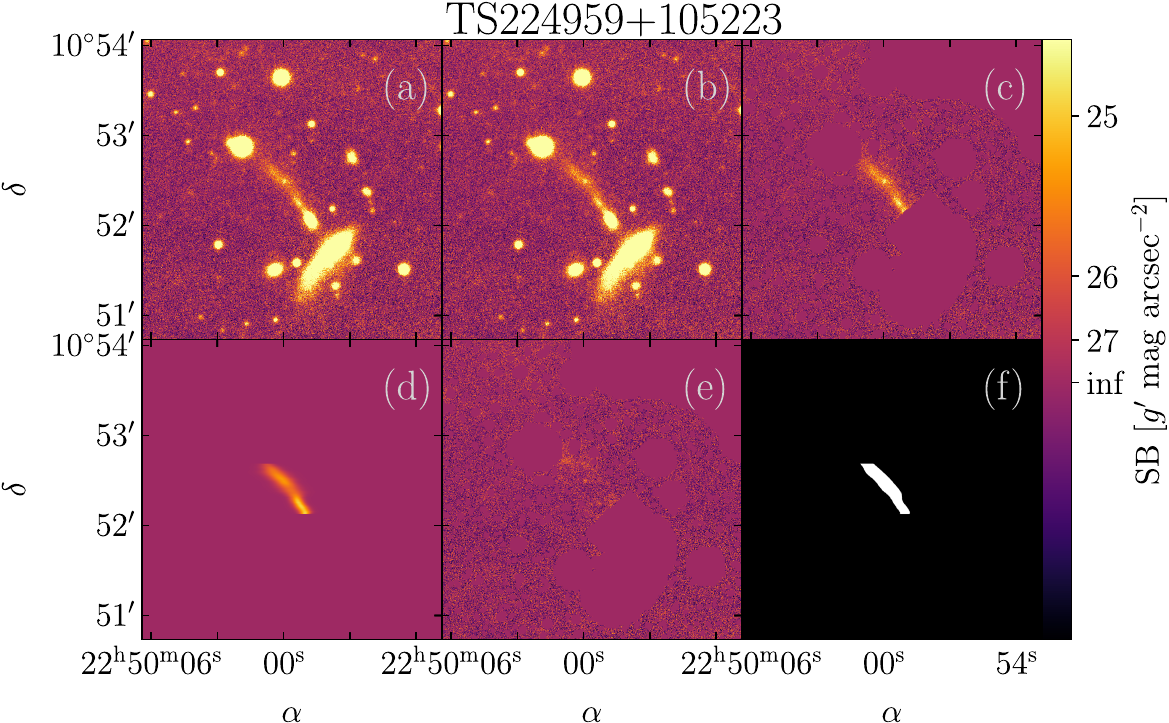}
    \caption{Same as in \hyperref[fig:chap-results:ts0120+1930]{Figure \ref{fig:chap-results:ts0120+1930}} but for the TS224959+105223 stream and without subtracting an isophotal model.}
\end{figure}
\begin{figure}
    \centering
    \includegraphics[width=\linewidth]{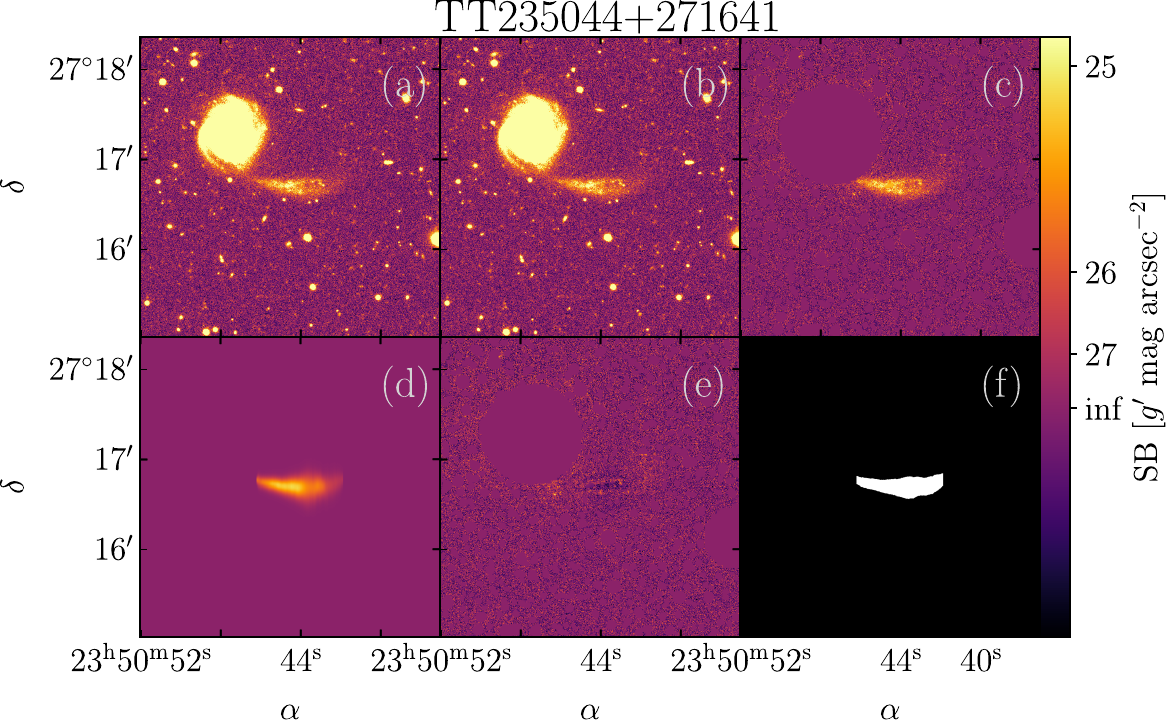}
    \caption{Same as in \hyperref[fig:chap-results:ts0120+1930]{Figure \ref{fig:chap-results:ts0120+1930}} but for the TT235044+271641 tail and without subtracting an isophotal model.}
\end{figure}
\begin{figure}
    \centering
    \begin{subfigure}
        \centering
        \includegraphics[width=\linewidth]{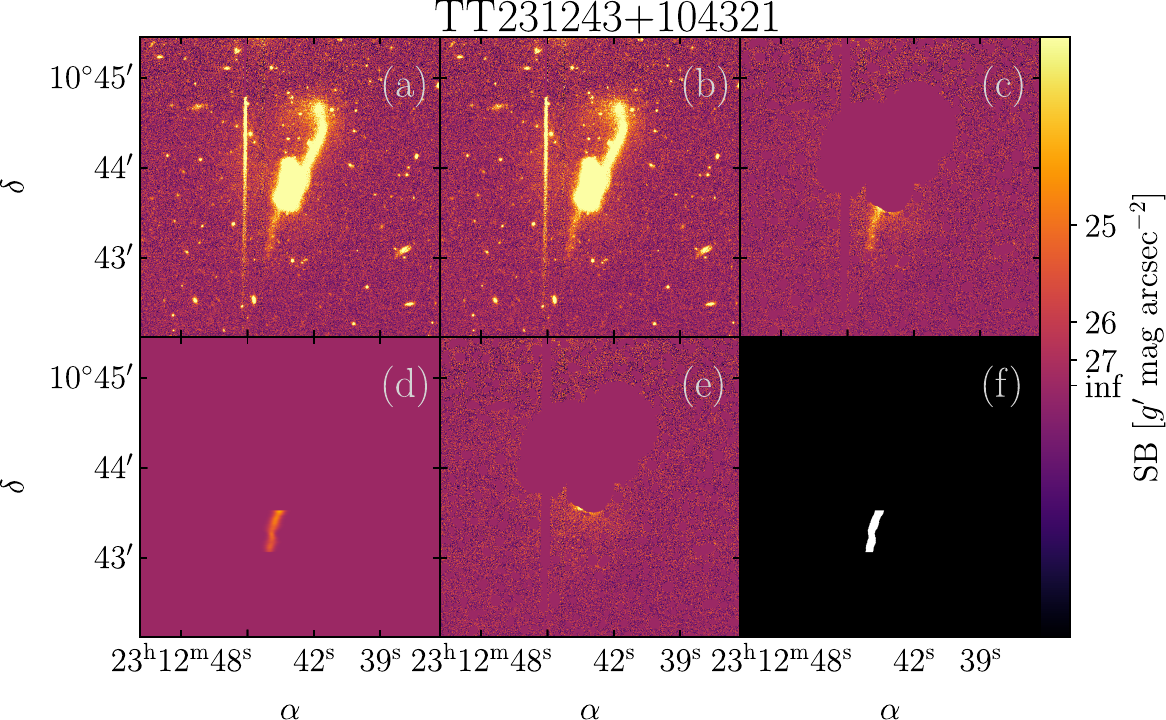}
    \end{subfigure}
    \begin{subfigure}
        \centering
        \includegraphics[width=\linewidth]{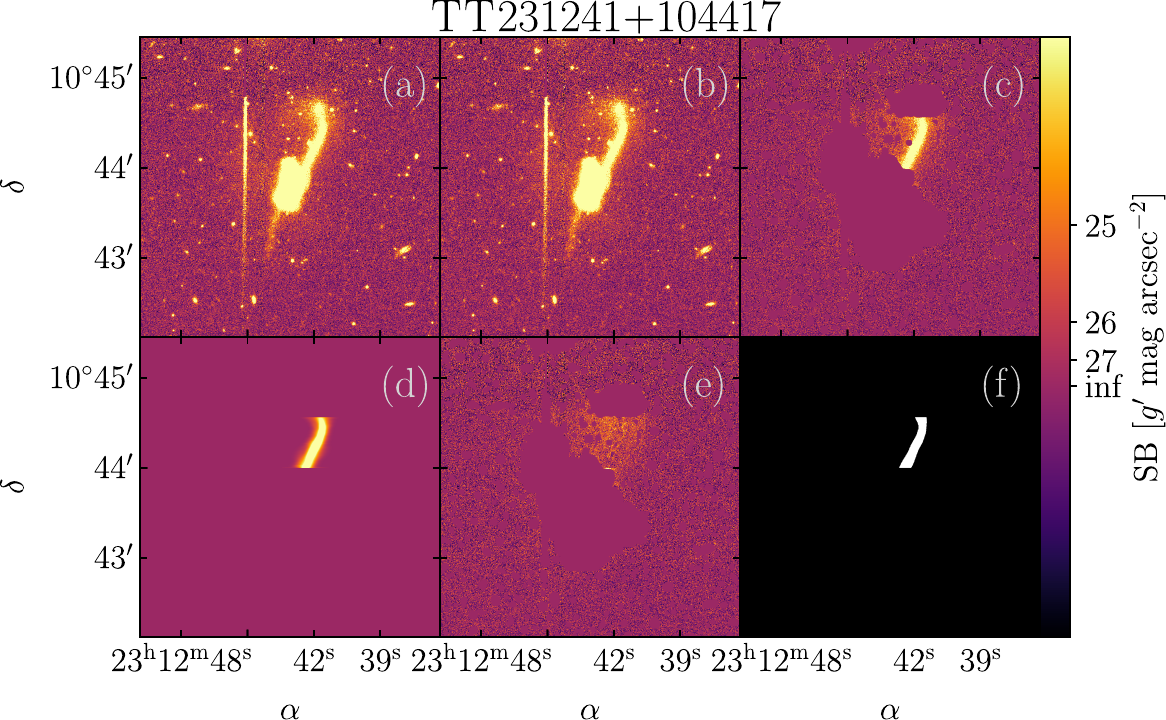}
    \end{subfigure}
    \caption{Same as in \hyperref[fig:chap-results:ts0120+1930]{Figure \ref{fig:chap-results:ts0120+1930}} but for the tails TT231243+104321 and TT231241+104417 and without subtracting an isophotal model.}
\end{figure}
\newpage
\begin{figure}
    \centering
    \begin{subfigure}
        \centering
        \includegraphics[width=\linewidth]{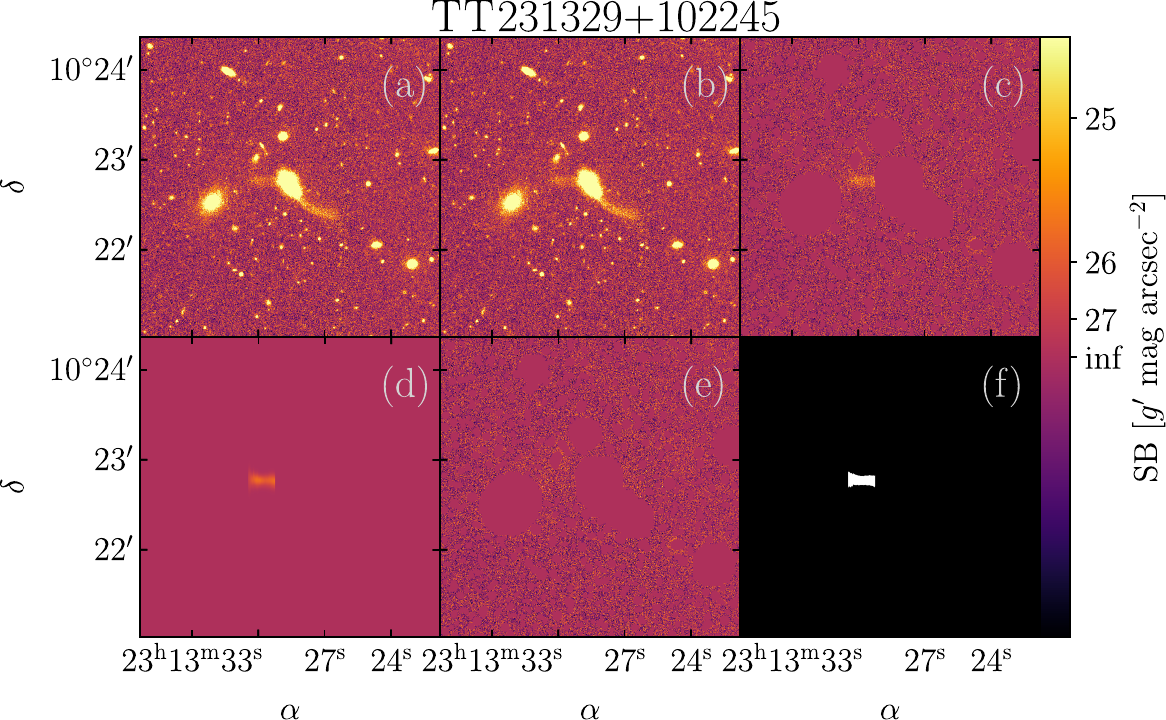}
    \end{subfigure}
    \begin{subfigure}
        \centering
        \includegraphics[width=\linewidth]{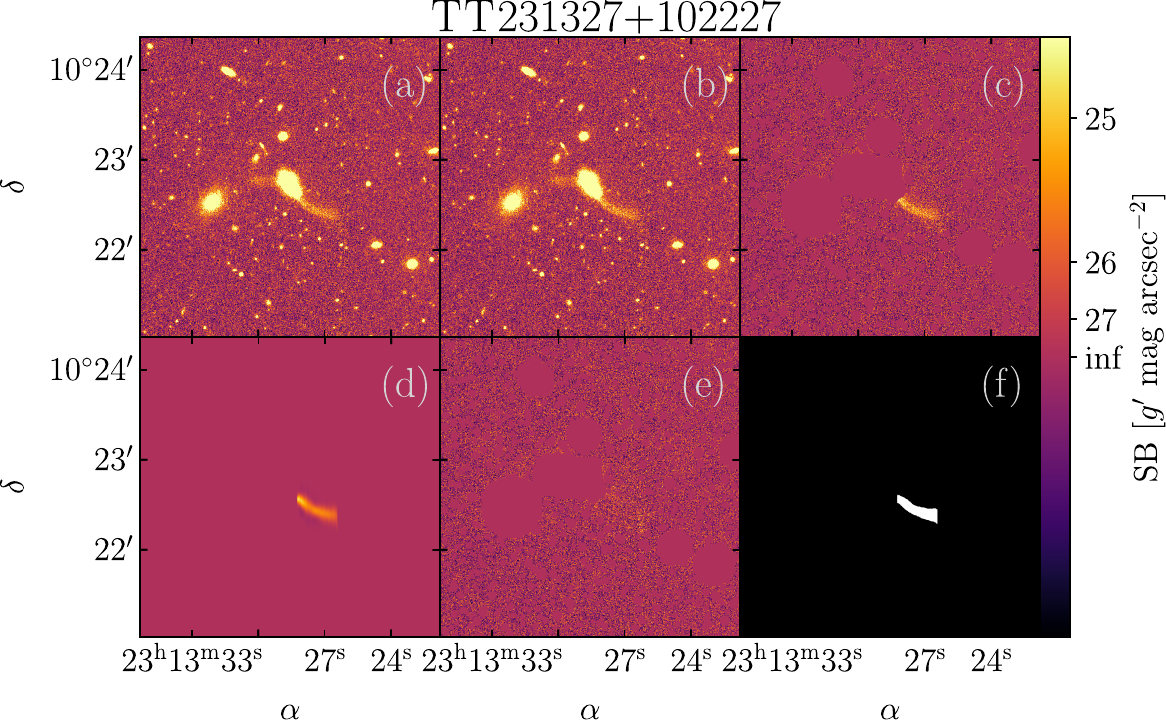}
    \end{subfigure}
    \caption{Same as in \hyperref[fig:chap-results:ts0120+1930]{Figure \ref{fig:chap-results:ts0120+1930}} but for the tails TT231329+102245 and TT231327+102227 and without subtracting an isophotal model.}
    \label{fig:app:streams:end}
\end{figure}
\section{Classified Tidal Features}
Table \ref{tab:allfeatures} and \ref{tab:allfeaturescont} contain all tidal features identified and classified during the WWFI Abell $g'$-band manual image inspection. The ones used in the analysis of this paper are not listed but found in Table \ref{tab:feature-list}. Features separated by '/' indicate an OR case, where '\&' means that both features are present in the image. The classified features were also cross checked with their appearance in the Legacy Survey DR10 to help with the classification using color information.
\begin{table*}
    \footnotesize
    \centering
    \begin{tabular}{c c c c c c}
        \hline
        \hline
        $\alpha_\text{Feature}$ & $\delta_\text{Feature}$ & $z_\text{Feature}$ & $z$ Candidate & Angular Scale & Feature \\
        J2000 & J2000 & & & [kpc/$''$] & \\
        (1) & (2) & (3) & (4) & (5) & (6) \\
        \hline
        17.878888 & 17.457995 & -         & 2MASX J01113078+1727282 & -     & Tail\&Stream \\
        17.882192 & 17.609381 & -         & 2MASX J01113133+1736353 & -     & Shell \\
        25.813381 &  6.554159 & - & - & - & Stream \\
        26.225615 &  6.382950 & 0.0918310 & LAMOST J014453.48+062301.0 & 1.721 & Stream/Tail \\
        27.283573 & 14.221135 & 0.1227400 & LEDA 1452449            & 2.221 & Stream/Tail \\
        27.285417 & 13.962861 & 0.0705500 & 2MASX J01490841+1357474 & 1.355 & Shell \\
        27.451133 & 13.846310 & 0.0438800 & 2MASX J01494842+1350450 & 0.869 & Stream \\
        44.154073 & 13.004558 & 0.0660620 & 2MASX J02563682+1300191 & 1.275 & Stream/Shell \\
        79.270523 &  6.172417 & 0.0423890 & UGC 3278 & 0.841 & Tail \\
        93.569523 & 48.378326 & - & - & - & Tail/Spiral Arm \\
        100.091940 & 69.967313 & - & LEDA 2733304 & - & Tail/Spiral Arm \\
        116.044925 &  9.234112 & 0.0609625 & 2MASX J07441076+0914036 & 1.184 & Shell \\
        117.355476 & 52.043750 & 0.0683000 & MCG+09-13-062 & 1.315 & Stream/Tail \\
        118.188519 & 29.405885 & 0.0549480 & NVSS J075245+292419 & 1.075 & Stream/Tail/Spiral Arm \\
        118.504914 & 29.420762 & 0.0360400 & LEDA 1865757 & 0.721 & Stream/Tail/Spiral Arm \\
        118.506025 & 29.437427 & 0.1022400 & LEDA 1866269 & 1.894 & Tail \\
        123.665965 & 58.222177 & 0.0268800 & UGC 4281 & 0.543 & Tail/Merger \\
        123.936703 & 58.321017 & 0.0270500 & UGC 4289 & 0.547 & Shell \\
        127.182836 & 30.405352 & 0.0526200 & IC 2380 & 1.032 & Shell \\
        127.422200 & 30.688373 & 0.0476200 & IC 2383 & 0.939 & Shell \\      
        136.910766 & 16.592621 & 0.0729300 & 2MASX J09073872+1635334 & 1.397 & Shell \\
        147.979809 &  5.712064 & 0.0732700 & LEDA 1287878 & 1.403 & Stream/Tail \\
        156.801606 & 10.984194 & 0.0337522 & 2MASX J10271185+1059573 & 0.677 & Tail \\
        157.091027 & 10.727597 & 0.2014100 & SDSSCGB 8207.1          & 3.344 & Tail \\        
        159.021312 & 41.421745 & 0.1198300 & LEDA 2180848            & 2.176 & Tail/Spiral Arm \\
        159.025891 & 41.453541 & 0.1202100 & 2MASS J10360603+4126022 & 2.182 & Stream \\
        159.502640 & 41.764194 & 0.1199130 & 2MASX J10380107+4145534 & 2.177 & Tail \\                      
        164.940379 & 10.796817 & 0.0386400 & LEDA 33115              & 0.770 & Stream/Tail \\
        165.188659 & 10.568083 & 0.0338400 & IC 664                  & 0.679 & Stream \\
        168.101120 & 39.613875 & 0.0750600 & LEDA 2152868            & 1.434 & Tail/Spiral Arm \\
        169.081513 & 29.253911 & 0.0452800 & NVSS J111622+291507     & 0.896 & Stream/Tail \\
        171.703864 & 35.249622 & 0.0322500 & Mrk 423                 & 0.648 & Tail \\
        172.297598 & 54.116694 & 0.0683700 & 2MASX J11291167+5407028 & 1.316 & Tail \\     
        175.461621 & 10.276678 & 0.1188600 & 2MASX J11415073+1016340 & 2.160 & Stream\&Tail \\
        175.520162 & 10.339691 & 0.0207400 & NGC 3819                & 0.422 & Stream \\ 
        175.529333 & 10.147528 & 0.1187900 & 2MASX J11420742+1008578 & 2.159 & Stream \\
        175.692387 & 10.301782 & 0.1179100 & SDSS J114246.57+101813.4& 2.145 & Tail/Ring Spiral Arm \\      
        176.225352 & 19.778392 & 0.0274200 & 2MASX J11445486+1946349 & 0.554 & Shell \\
        176.275089 & 19.794546 & 0.067*    & 2MASX J11450584+1947329 & 1.292 & Stream \\
        176.287225 & 19.812583 & 0.0684894 & [CGB2004b] J114509+194845 & 1.319 & Stream \\
        176.322840 & 20.021520 & 0.0177720 & 2MASX J11451755+2001106 & 0.363 & Merger \\
        177.778368 & 55.078017 & 0.0196800 & NGC 3921                & 0.401 & Tail \\
        178.516940 & 23.340012 & 0.0509300 & LEDA 37330              & 1.001 & Tail/Spiral Arm \\
        179.354138 & 33.678161 & 0.1261600 & SDSS J115724.81+334043.5& 2.274 & Tail/Spiral Arm \\
        180.735731 & 51.696392 & 0.0630886 & 2MASX J12025666+5141437 & 1.222 & Stream/Shell/Tail \\ 
        180.747194 & 51.827954 & 0.2863400 & NVSS J120259+514947     & 4.349 & Tail/Spiral Arm \\
        180.821093 & 51.402197 & 0.0296610 & Mrk 1465                & 0.598 & Merger \\
        184.954941 &  5.464035 & 0.0079605 & NGC 4270                & 0.165 & Shell \\
    \end{tabular}
    \caption{List of Tidal Features Found in the WWFI Abell cluster sample. All redshifts were determined with SIMBAD \citep{wenger+2000}. The R.A. and decl. are given in degrees.}
    \label{tab:allfeatures}
\end{table*}
\begin{table*}
    \footnotesize
    \centering
    \begin{tabular}{c c c c c c}
        \hline
        \hline
        $\alpha_\text{Feature}$ & $\delta_\text{Feature}$ & $z_\text{Feature}$ & $z$ Candidate & Angular Scale & Feature \\
        J2000 & J2000 & & & [kpc/$''$] & \\
        (1) & (2) & (3) & (4) & (5) & (6) \\
        \hline
        185.426955 & 13.755289 & 0.0825200 & LEDA 169111             & 1.563 & Stream/Tail \\    
        191.829224 & 30.020983 & 0.1386200 & SDSS J124719.29+300123.6& 2.465 & Stream/Merger \\
        195.233755 & 27.791371 & 0.0265900 & NGC 4911                & 0.538 & Shell \\
        197.758424 & 39.192192 & 0.0781500 & 2MASX J13110199+3911281 & 1.488 & Tail \\
        202.488028 & 37.295658 & 0.0554320 & 2MASX J13295738+3717447 & 1.084 & Tail \\     
        203.610585 & 59.374177 & 0.0721900 & SDSS J133432.04+592156.8& 1.384 & Stream \\
        204.035015 & 59.206639 & 0.0733812 & MCG+10-19-096           & 1.405 & Shell \\       
        207.360545 & 26.461400 & 0.0763800 & 2MASX J13492678+2627426 & 1.457 & Shell\&Stream \\
        207.421321 & 26.717214 & 0.0665700 & 3XMM J134941.2+264301   & 1.285 & Tail/Stream/Shell \\ 
        208.276422 &  5.151083 & 0.0788500 & 2MASX J13530637+0508586 & 1.500 & Shell \\      
        209.540299 & 18.341026 & 0.0631700 & 2MASX J13580938+1820217 & 1.224 & Tail/Spiral Arm \\
        209.672206 & 20.302733 & 0.064*    & LEDA 1623208            & 1.239 & Tail/Spiral Arm \\
        209.749308 & 28.068404 & 0.0628100 & LEDA 93770              & 1.217 & Stream/Spiral Arm \\
        215.258601 & 17.699847 & 0.0512600 & IC 1004                 & 1.007 & Shell \\
        215.375845 & 17.571912 & 0.0509200 & LEDA 1535967            & 1.001 & Tail \\
        216.535770 & 16.788718 & 0.0873100 & 2MASX J14260855+1647167 & 1.645 & Stream\&Tail \\     
        222.718011 & 30.524442 & 0.0608100 & LEDA 1907325            & 1.181 & Stream \\
        222.813522 & 30.474584 & 0.1629800 & 2MASX J14511539+3028213 & 2.821 & Tail \\
        227.746036 &  5.743472 & 0.0891800 & LEDA 54175              & 1.676 & Tail/Spiral Arm \\
        229.188233 &  7.176194 & 0.1033615 & 2MASX J15164535+0710330 & 1.912 & Tail\&Merger \\
        230.325925 & 30.833527 & 0.0740800 & 2MASX J15211835+3049584 & 1.417 & Shell/Spiral Arm \\
        230.593450 & 27.557305 & 0.0732700 & 2MASX J15222245+2733262 & 1.403 & Stream/Tail \\
        230.640788 & 27.644913 & 0.0638800 & 2MASX J15223389+2738441 & 1.236 & Shell \\        
        240.532971 & 15.697583 & 0.0344500 & IC 1165b      & 0.690 & Tail \\
        240.551719 & 15.908417 & 0.0357900 & MCG+03-41-051 & 0.716 & Shell \\     
        240.887740 & 16.323267 & 0.0381030 & MCG+03-41-061 & 0.760 & Shell \\
        241.247652 & 17.871129 & 0.0331900 & NGC 6044      & 0.666 & Shell \\
        241.388719 & 17.595926 & 0.0339800 & IC 1181       & 0.681 & Tail \\
        241.404705 & 17.802575 & 0.0352136 & IC 1182       & 0.705 & Shell\&Stream \\
        241.418981 & 18.057720 & 0.0400500 & LEDA 1548961  & 0.797 & Tail \\
        241.672458 & 16.320024 & 0.0367190 & UGC 10204     & 0.734 & Shell/Tail \\
        242.937131 & 29.748091 & 0.0508100 & LEDA 200332   & 0.999 & Tail \\
        242.991003 & 29.837605 & 0.0500680 & UGC 10262     & 0.985 & Shell \\
        255.702219 & 78.742003 & 0.0669963 & 2MASX J17024809+7844270 & 1.292 & Shell \\
        290.633032 & 44.059785 & 0.0493070 & 2MASX J19223199+4403385 & 0.971 & Stream \\
        340.051613 & 12.177075 & - & - & - & Stream \\
        340.113598 & 12.361590 & - & - & - & Stream/Ring \\       
        345.275320 & 18.879946 & 0.0341336 & 2MASX J23010596+1852475 & 0.684 & Tail \\      
        354.251329 & 20.651082 & 0.0578250 & SDSS J233658.90+203910.1 & 1.127 & Tail/Merger \\
        354.273366 & 15.933285 & 0.0660700 & LEDA 1496749 & 1.276 & Tail/Spiral Arm \\
        354.325872 & 15.938152 & 0.1467600 & SDSS J233717.91+155616.1 & 2.586 & Stream/Merger \\
        354.463773 & 20.961078 & 0.0700284 & LEDA 1639646           & 1.346 & Tail/Spiral Arm \\
        354.622946 & 27.033706 & 0.0303543 & NGC 7720 & 0.611 & Shell \\
        354.672522 & 15.955254 & 0.0133500 & NGC 7722               & 0.274 & Shell \\
        354.727457 & 21.489628 & 0.0551300 & LEDA 1651104 & 1.078 & Tail \\
        358.994947 & 11.615024 & 0.0707550 & LEDA 1396212 & 1.359 & Stream \\
        359.007999 & 11.688415 & 0.0720680 & LEDA 1397303 & 1.382 & Tail/Stream \\
        359.019950 & 11.066182 & 0.075*    & - & 1.433 & Stream/Tail \\
        \hline
    \end{tabular}
    \caption{Continued list of WWFI tidal features of Table \ref{tab:allfeatures}.}
    \label{tab:allfeaturescont}
\end{table*}


\end{document}